\documentclass[sigconf]{acmart}
\AtBeginDocument{%
  }

\copyrightyear{2025}
\acmYear{2025}
\setcopyright{cc}
\setcctype{by}
\acmConference[CIKM '25]{Proceedings of the 34th ACM International
Conference on Information and Knowledge Management}{November 10--14,
2025}{Seoul, Republic of Korea}
\acmBooktitle{Proceedings of the 34th ACM International Conference on
Information and Knowledge Management (CIKM '25), November 10--14, 2025,
Seoul, Republic of Korea}\acmDOI{10.1145/3746252.3761047}
\acmISBN{979-8-4007-2040-6/2025/11}

\usepackage{xcolor}
\usepackage{balance}
\usepackage{mathtools}
\usepackage{url}
\usepackage{amsfonts}
\usepackage{algorithmic}
\usepackage{textcomp}
\usepackage{algorithm}
\usepackage{multirow}
\usepackage{color}
\usepackage{subcaption}
\usepackage{enumitem}

\usepackage{graphicx, caption, subcaption}

\usepackage{pgfplots}
\pgfplotsset{compat=1.17}

\usepackage{pdfrender}

\newcommand{\method}{\textsf{LIME}}
\newcommand{\codeurl}{\url{https://github.com/seongeunryu/lime-cikm25}}

\begin{document}


\title[Is This News Still Interesting to You?: Lifetime-aware Interest Matching for News Recommendation]{Is This News Still Interesting to You?:\\Lifetime-aware Interest Matching for News Recommendation}


\author{Seongeun Ryu}
\affiliation{%
  \institution{Hanyang University}
  \city{Seoul}
  \country{Republic of Korea}
}
\email{ryuseong@hanyang.ac.kr}

\author{Yunyong Ko}
\affiliation{%
  \institution{Chung-Ang University}
  \city{Seoul} 
  \country{Republic of Korea}
}
\email{yyko@cau.ac.kr}

\author{Sang-Wook Kim}
\authornote{Corresponding author}
\affiliation{%
  \institution{Hanyang University}
  \city{Seoul}
  \country{Republic of Korea}
}
\email{wook@hanyang.ac.kr}



\renewcommand{\shortauthors}{Seongeun Ryu, Yunyong Ko, and Sang-Wook Kim}

\begin{abstract}

Personalized news recommendation aims to deliver news articles aligned with users' interests, serving as a key solution to alleviate the problem of information overload on online news platforms. While prior work has improved interest matching through refined representations of news and users, the following time-related challenges remain underexplored: (\textbf{C1}) \textit{leveraging the age of clicked news to infer users’ interest persistence}, and (\textbf{C2}) \textit{modeling the varying lifetime of news across topics and users}. To jointly address these challenges, we propose a novel \underline{L}ifetime-aware \underline{I}nterest \underline{M}atching framework for n\underline{E}ws recommendation, named \textbf{{\method}}, which incorporates three key strategies: (1) User-Topic lifetime-aware age representation to capture the relative age of news with respect to a user-topic pair, (2) Candidate-aware lifetime attention for generating temporally aligned user representation, and (3) Freshness-guided interest refinement for prioritizing valid candidate news at prediction time. Extensive experiments on two real-world datasets demonstrate that {\method} consistently outperforms a wide range of state-of-the-art news recommendation methods, and its model-agnostic strategies significantly improve recommendation accuracy.

\end{abstract}



\maketitle

\section{Introduction}\label{sec:intro}

Personalized news recommendation aims to provide news articles that align with the interest of an individual user, 
which serves as a pivotal role for alleviating information overload on online news platforms.
A general approach in existing works~\cite{wu2019npa, wu2019nrms, wu2019naml, an2019lstur, wu2019tanr, wang2020fim, liu2020hypernews, qi2021hierec, mao2021cnesue, mao2022digat, chen2023tccm, yang2023glory, li2022miner, ko2025crown} typically follows a three-step pipeline: 
(1) (\textbf{news representation}) a news article is represented based on its attributes (e.g., title, body, topic) as a news embedding, 
(2) (\textbf{user representation}) the embeddings of historically clicked news of a user are aggregated into a user embedding, 
and (3) (\textbf{interest matching}) the degree of matching between the user and candidate news embeddings is estimated as an interest score. 
Accordingly, prior works on personalized recommendations have focused on enhancing news modeling~\cite{wu2019npa, wu2019nrms, wu2019naml, wu2019tanr, wang2020fim, kim2022cast, liu2020hypernews, mao2021cnesue, mao2022digat, yang2023glory, ko2025crown}, user modeling~\cite{wu2019npa,wu2019nrms, an2019lstur, mao2021cnesue, mao2022digat, ko2025crown}, or interest matching~\cite{qi2021hierec, chen2023tccm, qi2021pprec, qi2021kim, li2022miner}.

\begin{figure}[t]
\vspace{3mm}
\centering
\includegraphics[width=0.95\linewidth]{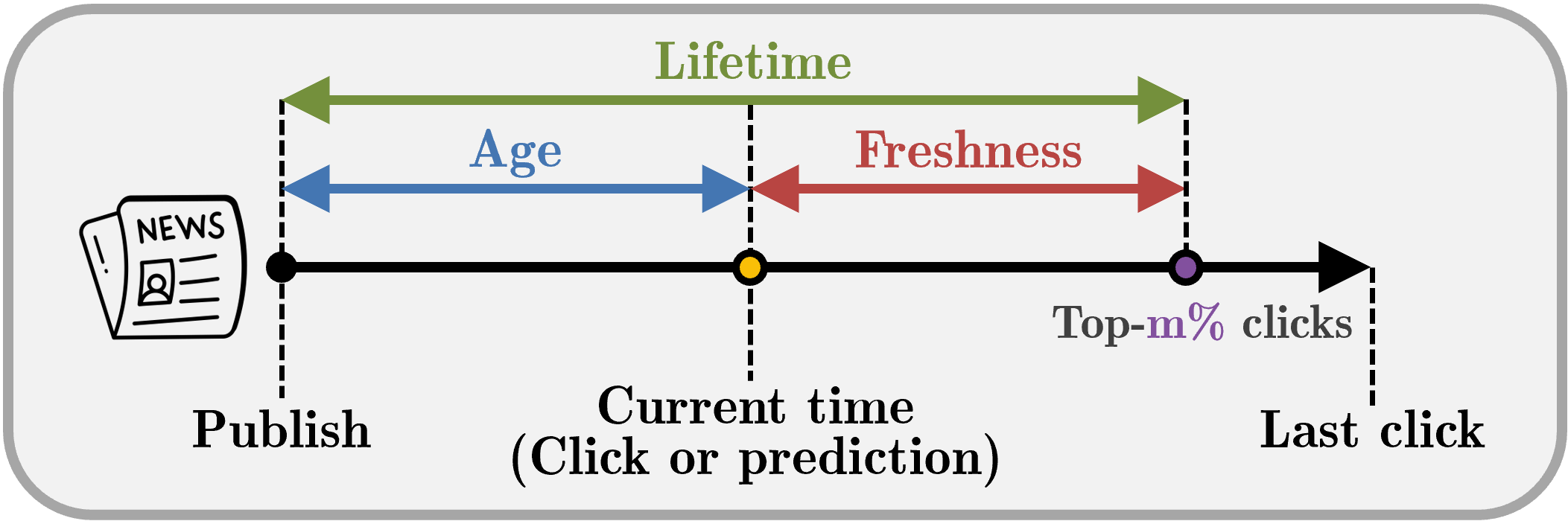}
\vspace{-3mm}
\caption{Age, lifetime, and freshness of news.}
\vspace{-5mm}
\label{fig:figure1}
\end{figure}

\begin{figure}[t]
\vspace{3mm}
\centering
\includegraphics[width=0.95\linewidth]{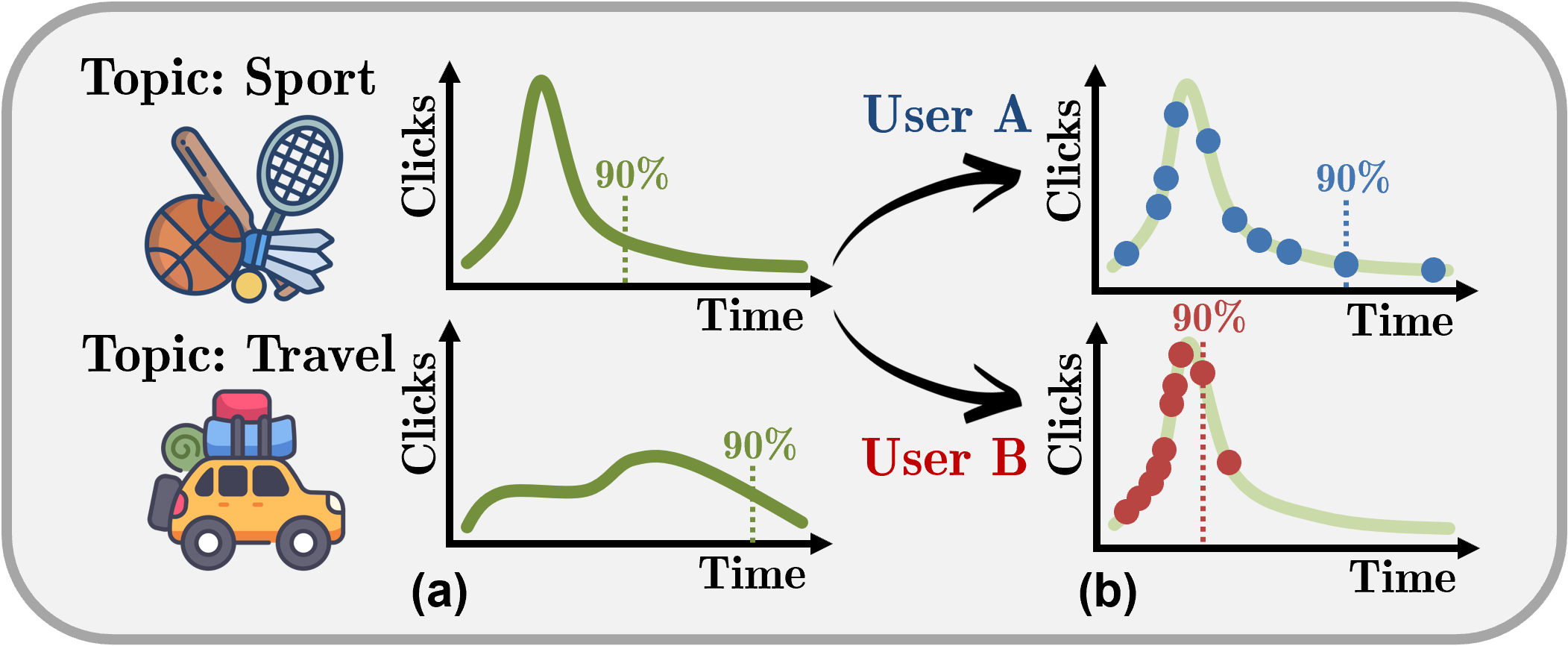}
\vspace{-3mm}
\caption{Topic-wise and user-wise news lifetimes.}
\vspace{-3mm}
\label{fig:figure2}
\end{figure}

\vspace{1mm}
\noindent
\textbf{Time-sensitive nature of the news domain.}
Compared to other recommendation domains, 
the news domain is inherently more \textit{time-sensitive}~\cite{wu2020mind, liu2020hypernews, wu2023personalized, qi2021pprec, chen2023tccm, bae2023lancer}. 
For instance, approximately 80\% of news articles in the MIND dataset~\cite{wu2020mind} received their last click within 48 hours of publication~\cite{wu2020mind}. 
In contrast, items in the Netflix dataset\footnote{\url{https://www.kaggle.com/datasets/netflix-inc/netflix-prize-data}} continued to receive interactions over several months, 
with 80\% of total interactions spread over an average of 32 months~\cite{bae2023lancer}.
This phenomenon indicates that the news content (e.g., title, body) \textit{quickly loses its validity to users over time}, 
which distinguishes the news domain from others. 


To address this time-sensitive nature of the news domain, 
recent studies~\cite{liu2020hypernews, qi2021pprec, chen2023tccm, bae2023lancer, cho2021overlooked} have attempted to incorporate temporal attributes, such as \textit{age}, \textit{freshness}, and \textit{lifetime} of news articles (see Figure~\ref{fig:figure1}), into news recommendation.
For example, HyperNews~\cite{liu2020hypernews} aims to capture the temporal dynamics of a news article by incorporating its \textit{age} (i.e., elapsed time from publish time) information into the news content embedding.
LANCER~\cite{bae2023lancer} introduces the concept of \textit{news lifetime}, defined as the elapsed time from publish time to the point at which a predefined top-$m\%$ (e.g., $m$=90) of clicks have occurred.

\vspace{1mm}
\noindent
\textbf{Challenges}.
Although existing works considering the time-sensitive nature of news have achieved higher accuracy, 
the following challenges still remain under-explored:

\textbf{(C1) Leveraging age of news in click history.} 
Intuitively, a user's click implies that the clicked news is \textit{valid} to her at that time.
This indicates that \textit{the age of clicked news in a user's click history contributes to better understanding her interest in a temporal perspective}.
Existing methods~\cite{liu2020hypernews, qi2021pprec, chen2023tccm, bae2023lancer}, however, have focused only on \textit{the age of candidate news} in order to filter out outdated news from recommendation, rather than user modeling,
which limits the potential to capture users' interests accurately in a temporal aspect.

\textbf{(C2) Varying lifetime across topics and users.}  
It is known that news articles have shorter lifetime (e.g., 36 or 48 hours) than those of items in other recommendation domains~\cite{bae2023lancer}.
Looking more closely, the lifetime of news content may vary \textit{across topics}.
For example, sport news about game results is quickly consumed and expires within a few hours, 
whereas travel news about landmarks or tips tends to remain valid for several days or even weeks (See Figure~\ref{fig:figure2}(a)).
Moreover, even for the same topic, users may differ in durations in which they remain interested.
For instance, a user \textit{A} may consume preferable news articles quickly,
while a user \textit{B} may consume them for a longer period (See Figure~\ref{fig:figure2}(b)).
Although one existing work~\cite{bae2023lancer} considers lifetime of news articles, it just adopts a \textit{single fixed lifetime} for all news, failing to exploit its full potential in news recommendation.

In summary, existing methods have the following limitations:
(1) \textit{they may recommend news articles that align with a user's interest but are no longer valid at that time}, 
and (2) \textit{they fail to capture a user's temporal aspect of interest and thus exclude preferable news from the recommendation}.
We will discuss these limitations in Section~\ref{sec:motivation}.

\vspace{1mm}
\noindent
\textbf{Our Approach.}
To address these limitations, in this paper, we first define two types of lifetime:  
\textit{Topic-wise lifetime} and \textit{User-Topic-wise lifetime} (\textit{User-Topic lifetime} in short) (See details in Section~\ref{sec:motivation}),
which serve as a foundation for better understanding the time-sensitive nature of news.
On the top of the definitions,
we propose a novel framework for personalized news recommendation, named \textbf{{\method}} (\underline{\textbf{L}}ifetime-aware \underline{\textbf{I}}nterest \underline{\textbf{M}}atching for n\underline{\textbf{E}}ws recommendation).
{\method} incorporates \textit{age}, \textit{lifetime}, and \textit{freshness} into news recommendation 
to effectively capture users' interest in a temporal aspect, thereby enhancing the interest matching.
It employs three key strategies:
(1) \textbf{User-Topic lifetime-aware age representation} (\textit{for news modeling}): to capture the relative age of news with respect to a user-topic pair;
(2) \textbf{Candidate-aware lifetime attention} (\textit{for user modeling}): 
to compute an attention weight for a clicked news to a candidate news based on their relative ages;
and (3) \textbf{Freshness-guided interest refinement} (\textit{for interest matching}): 
to estimate the final matching score by assessing whether the candidate news is still valid to a specific user at the time of recommendation.

The first two components incorporate temporal signals into user and news representations, 
and the third component leverages them to prioritize temporally relevant candidate news at the inference time.
To the best of our knowledge, {\method} is the first framework to jointly incorporate time-sensitive dynamics across both \textit{representation} and \textit{matching} levels.
It is worth noting that {\method} is \textit{model-agnostic} so that it can be integrated with any existing news recommendation backbone without structural modification. We show in Section~\ref{sec:eval-result-eq1} that {\method} helps improve the accuracy of state-of-the-art news recommendation models consistently.

\vspace{1mm}
\noindent
\textbf{Contributions.} 
Our main contributions are summarized as follows:
\begin{itemize}[leftmargin=10pt]
    \item \textbf{Observations and Definitions.} We conduct extensive analyses of news lifetime: based on the findings, we define fine-grained lifetime in terms of topic- and user-wise perspectives, which serves as a foundation for better understanding the time-sensitive nature of news recommendation.
    
    \item \textbf{Framework.} We propose {\method}, a lifetime-aware interest matching framework that employs (1) User-Topic lifetime-aware age representation, (2) Candidate-aware lifetime attention, and (3) Freshness-guided interest refinement in a unified and model-agnostic design.
    
    \item \textbf{Evaluation.} Through extensive experiments on two real-world datasets, we demonstrate that {\method} consistently outperforms existing state-of-the-art approaches, and that each of our proposed components is effective and contributes significantly to the overall performance.
\end{itemize}
\noindent
For reproducibility, we have released the code of {\method} and the datasets at {\codeurl}.

\begin{figure}[t]
\centering
\hspace*{-4mm}\includegraphics[width=0.95\linewidth]{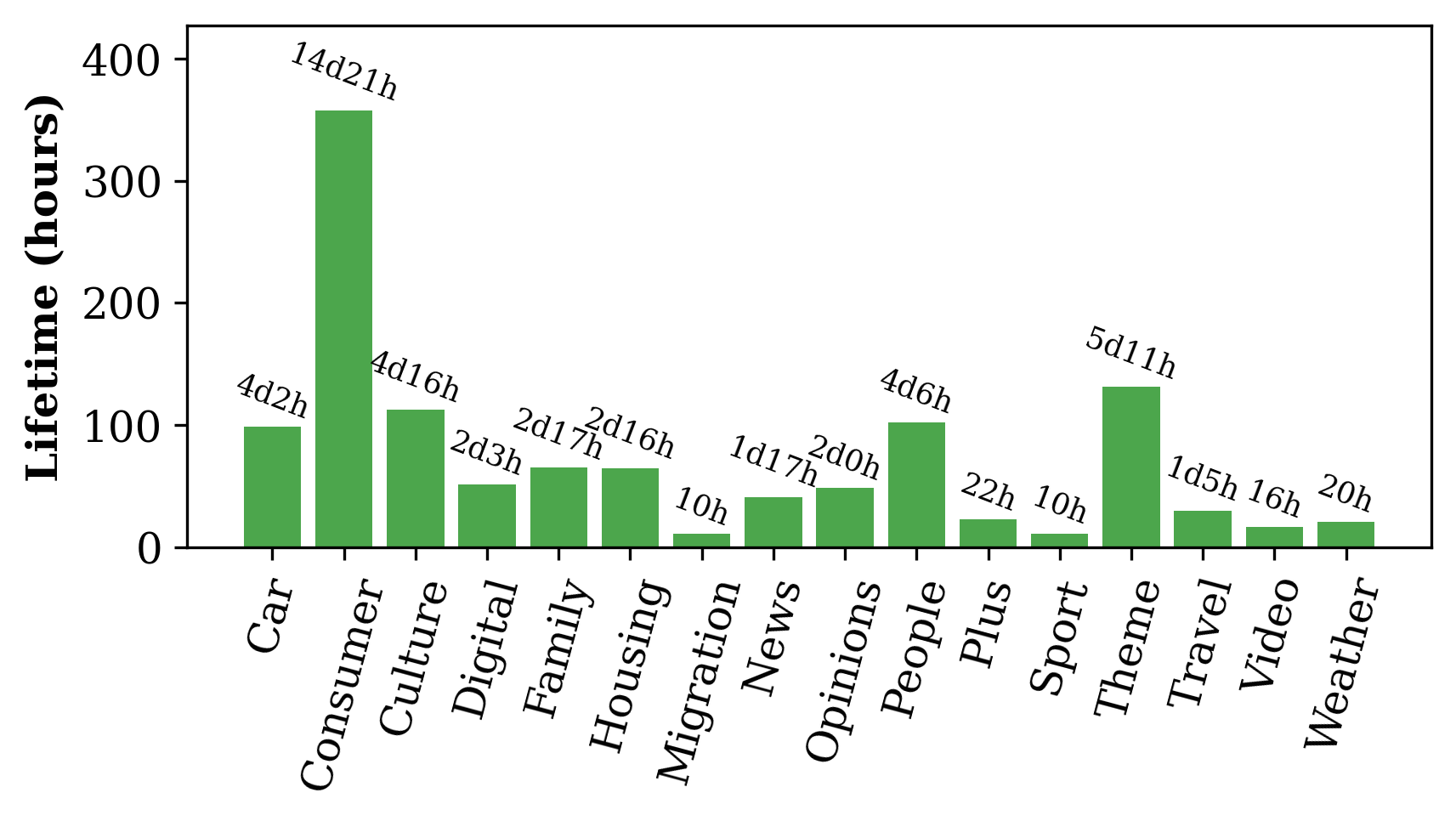} \\
\vspace{-1mm}
\textbf{(a) Topic-wise differences in news lifetime} \\
\vspace{3mm}
\includegraphics[width=1.00\linewidth]{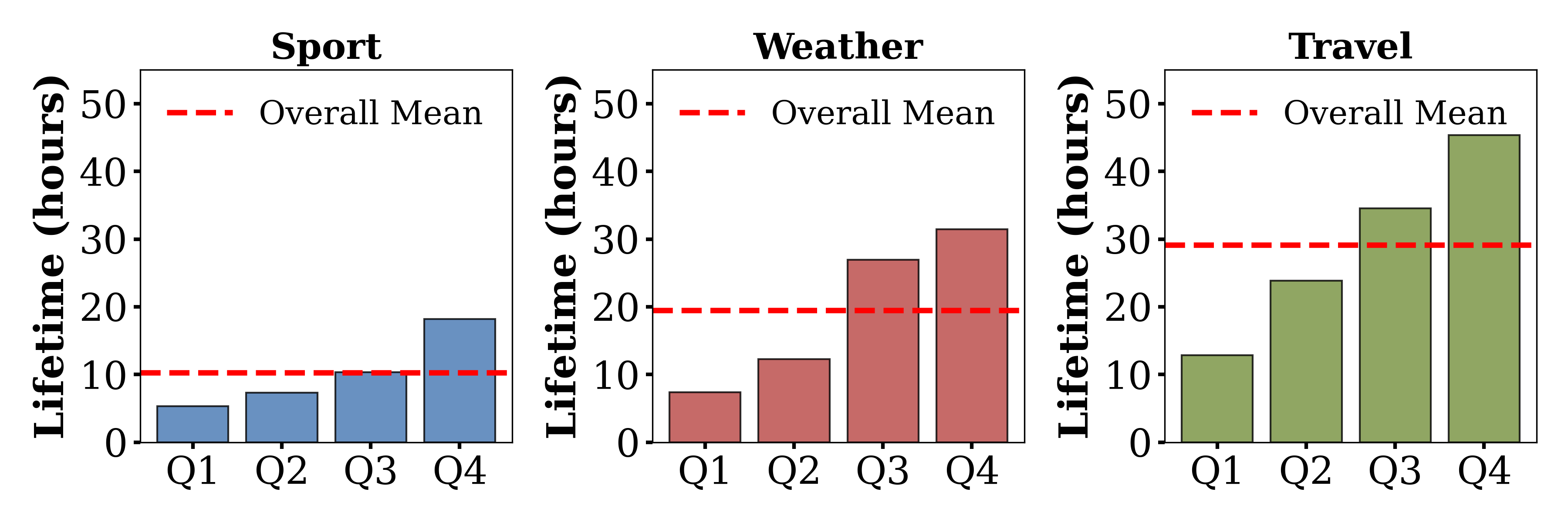} \\
\vspace{1mm}
\textbf{(b) User-wise differences in news lifetime within topics}
\vspace{-2mm}
\caption{Two key observations motivating {\method}: 
(a) Different news lifetimes across topics; 
(b) Different news lifetimes across user quartiles within the same topic.}
\vspace{-2mm}
\label{fig:figure3}
\end{figure}

\section{Motivation}
\label{sec:motivation}
We posit that news lifetime may vary depending on \textit{topics} and \textit{users}.
In this section, we conduct a preliminary analysis to verify our hypothesis.

\vspace{1mm}
\noindent
\textbf{(Observation 1) Topic-wise lifetime.}  
We categorize clicked news articles in the Adressa dataset~\cite{gulla2017adressa} by topic\footnote{Among 108 total topics in Adressa, we selected the top-16 topics with the highest number of user clicks.} and compute the average lifetime of news (i.e., the time until top-$90\%$ of clicks).
Figure~\ref{fig:figure3}(a) shows that \textit{significant differences in average lifetime across topics}.
For instance, \textit{Consumer} news remains valid for more than two weeks, while \textit{Sport} news is valid for about 10 hours, reflecting its quickly expiring content.

\vspace{1mm}
\noindent
\textbf{(Observation 2) User-wise lifetime.}  
We analyze users who clicked at least 10 articles from each of the three topics with distinct lifetimes—\textit{Sport}, \textit{Weather}, and \textit{Travel}.
We group the users into quartiles (i.e., Q1-Q4) based on their individual lifetime distributions and compare average lifetimes within each quartile.
Figure~\ref{fig:figure3}(b) shows that the average lifetime differs significantly across user groups.
For example, in the \textit{Travel} news, users in Q4 show average lifetime more than four times longer than those in Q1, 
indicating that the perceived validity period varies across users. We performed a \textit{t}-test between Q1 and Q4 groups and confirmed that the difference in average lifetime was statistically significant (\textit{p}-value < 0.001).

These observations suggest that \textit{the lifetime of news is not only topic-dependent but also user-dependent}.
Motivated by this, 
we define new concepts of topic-wise and user-topic lifetimes as follows:



\vspace{1mm}
\noindent
\textbf{Definition 1 (Topic-wise lifetime)}
\textit{The news lifetime $L(t, m)$ on topic $t$ is defined as the average time it takes for top-$m\%$ of clicks to occur across all news articles in $t$, measured up to their last click time.}

\vspace{1mm}
\noindent
\textbf{Definition 2 (User-Topic lifetime)}
\textit{The user-specific news lifetime $L(u, t, m)$ for user $u$ on topic $t$ is defined as the time from publication to the point at which top-$m\%$ of user $u$'s clicks on topic $t$ occur, measured up to their last click.}

\begin{figure}[t]
    \centering
    \begin{minipage}[b]{0.48\linewidth}
        \centering
        \includegraphics[width=\linewidth]{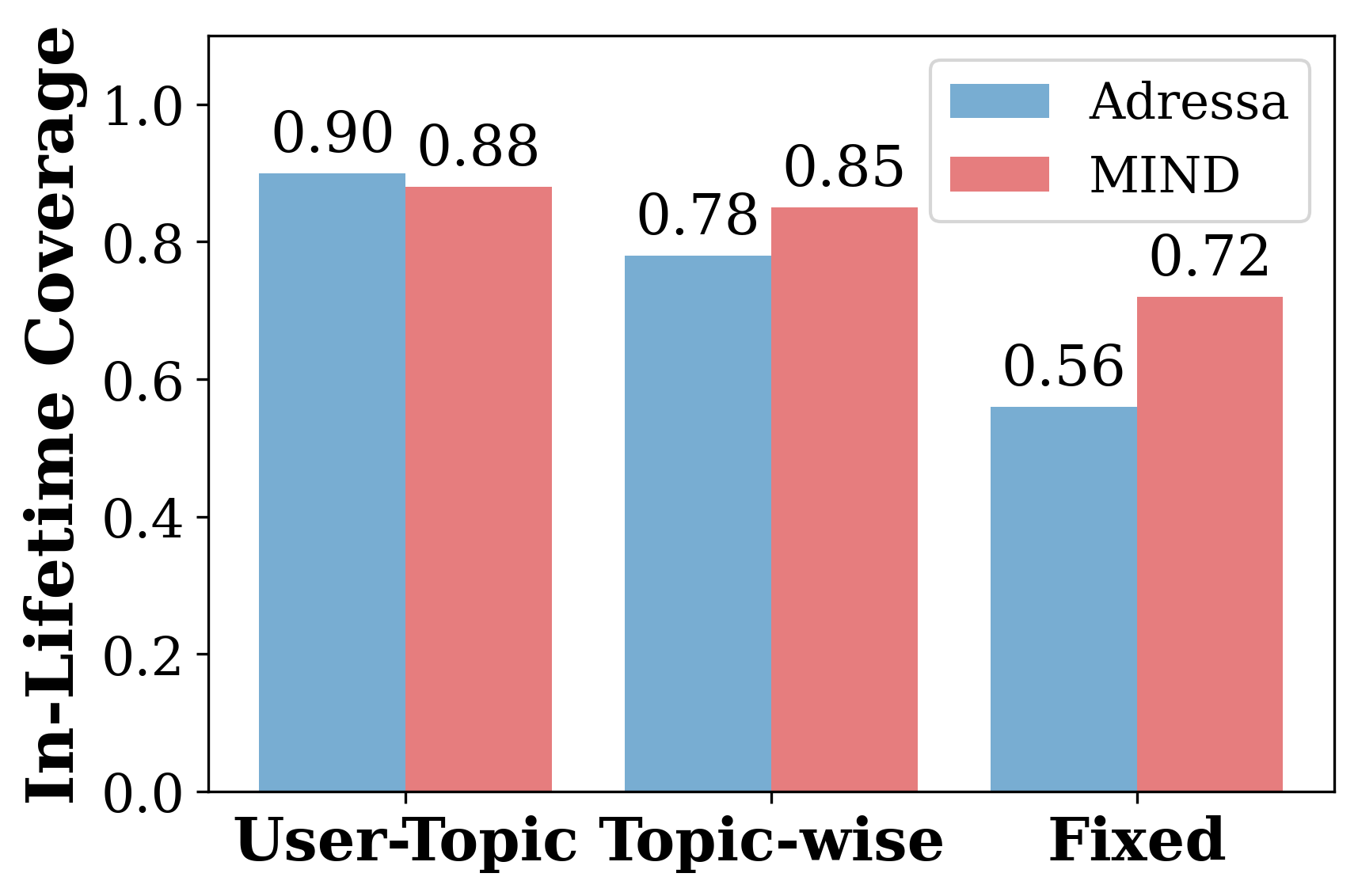}
        \subcaption{Positive click coverage}
    \end{minipage}
    \hfill
    \begin{minipage}[b]{0.48\linewidth}
        \centering
        \includegraphics[width=\linewidth]{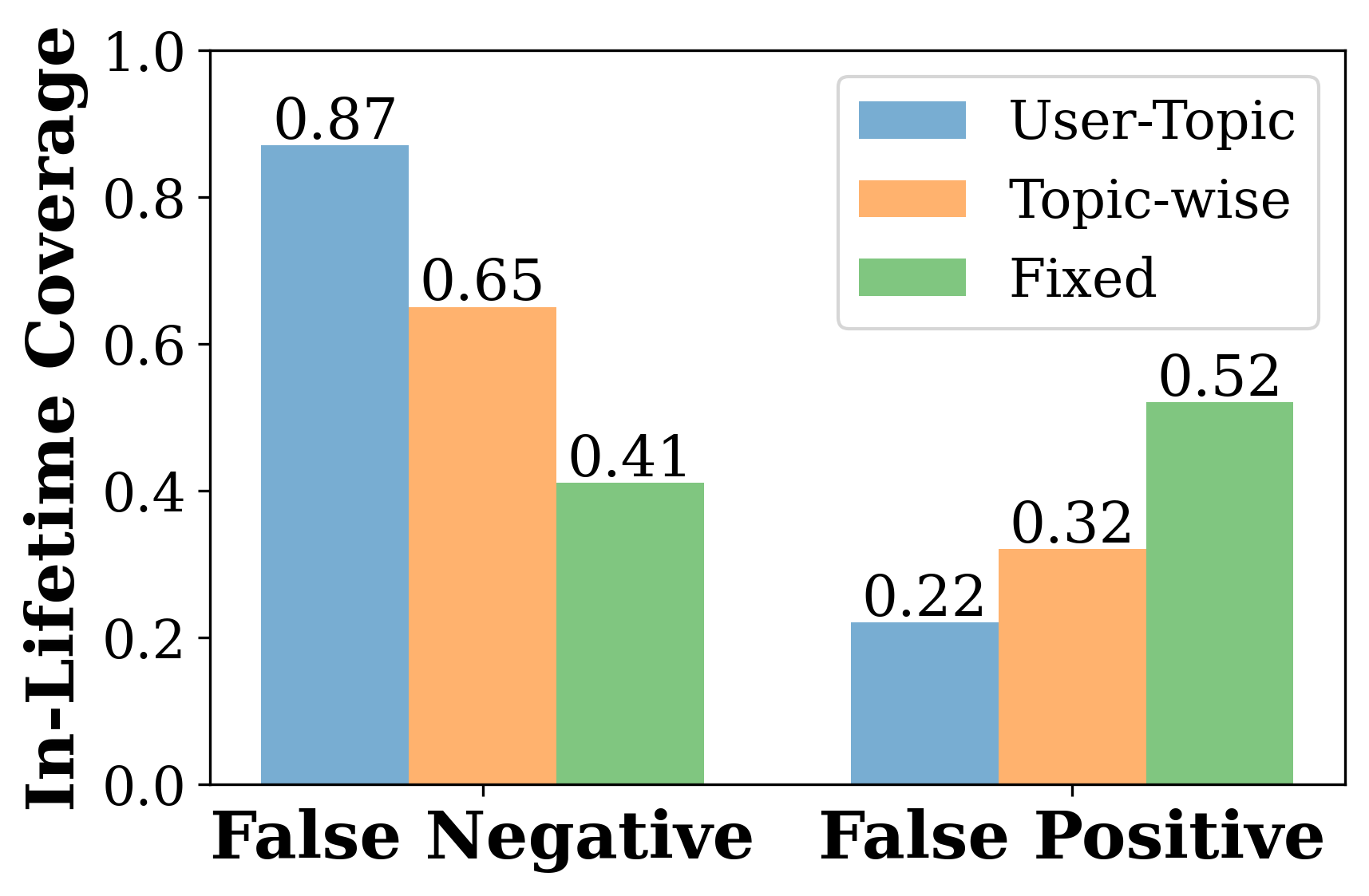}
        \subcaption{Prediction error coverage}
    \end{minipage}
    \vspace{-2mm}
    \caption{Coverage analysis across lifetime definitions: (a) Positive click; (b) Prediction error (in Adressa~\cite{gulla2017adressa}).}
    \label{fig:figure4}
\end{figure}

We also conduct two preliminary experiments to show the potential utility of our user-topic lifetime definition:

\vspace{1mm}
\noindent
\textbf{(Validation 1) Alignment with real-world click behaviors.}  
We examine how well each lifetime definition aligns with users' actual click behaviors under the assumption that users click a news article if they perceive it as still valid.
We compare the proportion of clicked news articles that fall within their respective lifetimes using three settings:
(1) Fixed lifetime~\cite{bae2023lancer},
(2) Topic-wise lifetime (\textit{ours}), and
(3) User-Topic lifetime (\textit{ours}).
As shown in Figure~\ref{fig:figure4}(a), the in-lifetime coverage of clicks is highest under user-topic lifetime (90.08\%), 
followed by topic-wise lifetime (78.05\%), and fixed lifetime (55.88\%).
This result indicates that the proposed user-topic lifetime aligns most closely with how users perceive the validity of content.  
These findings empirically demonstrate that user-topic lifetime is the most-realistic and behaviorally grounded definition, as it best aligns with actual user click behavior in the real world.

\vspace{1mm}
\noindent
\textbf{(Validation 2) Prediction errors in existing models.}  
We further evaluate whether user-topic lifetime can help explain prediction errors in existing models such as CROWN~\cite{ko2025crown}.
Specifically, we analyze whether false negative (FN) and false positive (FP) cases obtained from CROWN fall within the lifetime period, 
comparing the three lifetime definitions in Adressa\cite{gulla2017adressa}.
As shown in Figure~\ref{fig:figure4}(b), 87.25\% of FN cases are within the user-topic lifetime, 
suggesting that many of clicked articles not predicted by the model are still valid to the user.
In contrast, only 22.42\% of FP cases are within the user-topic lifetime, indicating that many of incorrectly predicted articles are those no longer valid to the user.

These results demonstrate that the \textit{user-topic lifetime not only aligns with real-world click behavior, but also has strong potential to reduce prediction errors in existing news recommendation models}.
Taken together, our findings highlight the limitations of fixed-lifetime modeling and motivate the need for a new framework that incorporates lifetime from both topic and user perspectives.

\begin{figure*}[t]
\centering
\includegraphics[width=0.95\textwidth]{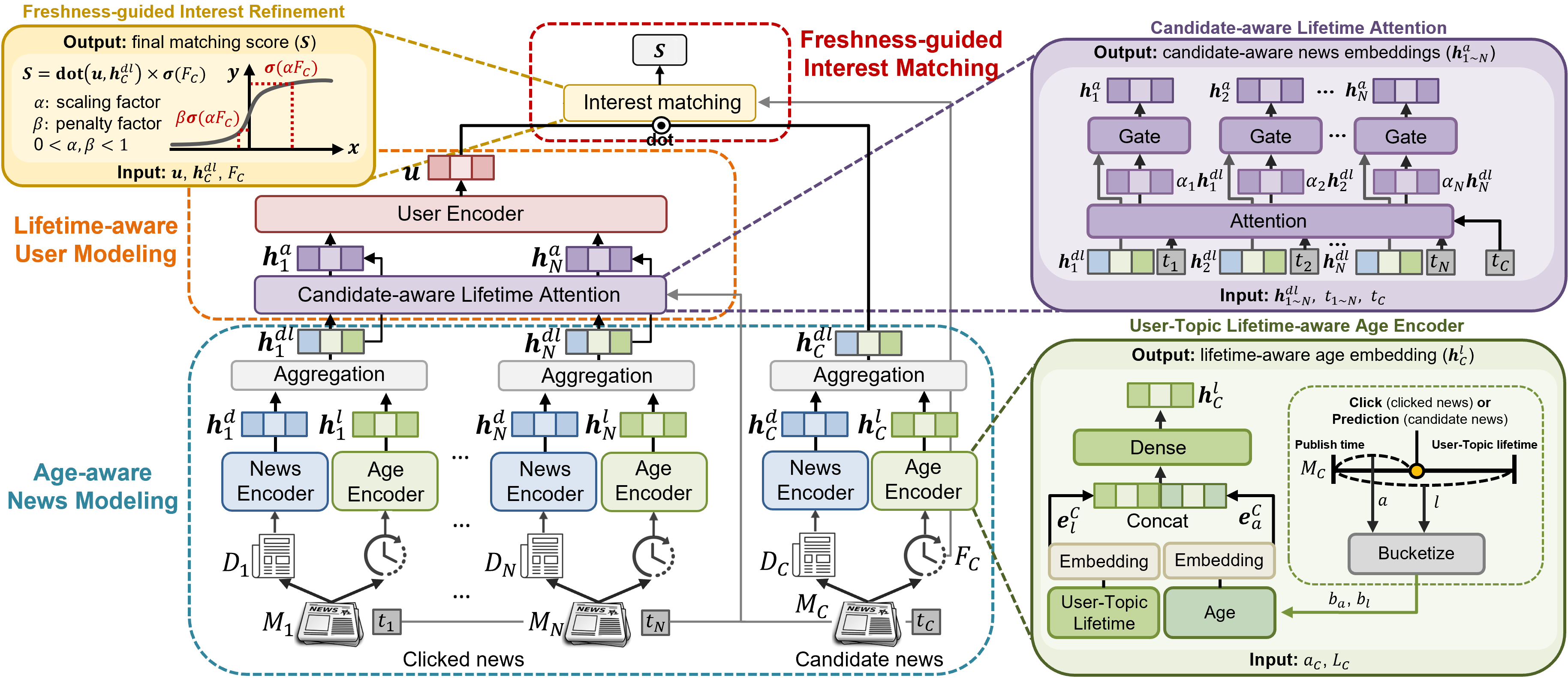}
\vspace{-3mm}
\caption{Overview of {\method}, which consists of three core components: age-aware news modeling, lifetime-aware user modeling, and freshness-guided interest matching.}
\vspace{-4mm}
\label{fig:overview}
\end{figure*}

\section{Proposed Framework: {\method}}\label{sec:proposed}

\noindent
In this section, we propose a novel model-agnostic framework, named \underline{\textbf{L}}ifetime-aware \underline{\textbf{I}}nterest \underline{\textbf{M}}atching for n\underline{\textbf{E}}ws recommendation (\textbf{{\method}}), 
which is designed to model the time-sensitive nature of news and user interest in a temporal perspective.

\subsection{Problem Definition}\label{sec:problem-definition}

\noindent
We define the task of personalized news recommendation as predicting whether a user will click a given candidate news article. 

Let $u$ be a user and $\mathcal{N}_u = [M_1, M_2, ..., M_N]$ be the sequence of $N$ news articles previously clicked by $u$,
with corresponding topics $\mathcal{T}_u = [t_1, t_2, ..., t_N]$, respectively. 
Each news article $M_n$ is associated with content attributes $D_n$, i.e., the title, body, and topic (which we use interchangeably with a category), and time-related attributes $L_n$, i.e., age, lifetime, freshness. 
Given a candidate news article $M_c$ with content $D_c$, temporal attributes $L_c$ (including freshness $F_c$), 
where its topic is denoted by $t_c$, the goal is to estimate the probability $\hat{y} \in \{0, 1\}$ that user $u$ will click on $M_c$.

In our setting, we further assume that each user-news pair is associated with two temporal attributes: 
\textit{age} (see Figure~\ref{fig:figure1}), 
and \textit{user-topic lifetime}, both of which can be used to refine the modeling of user interest and news validity over time. The prediction function $g(u, M_c) \rightarrow \hat{y}$ is learned based on enriched user and news representations, 
and final recommendations are made by ranking candidate news articles based on the predicted matching scores.

\subsection{Model Overview}\label{sec:model-overview}

\noindent
As illustrated in Figure~\ref{fig:overview}, {\method} consists of three key components: 
\textbf{age-aware news modeling}, \textbf{lifetime-aware user modeling}, and \textbf{freshness-guided interest matching}.

\begin{itemize}[leftmargin=10pt]
    \item \textbf{Age-aware news modeling} encodes each news article with a \textit{user-topic lifetime-aware age encoder} that captures how early or late the article is consumed compared to how long the user typically stays interested in the topic. The resulting age embedding is then combined with a content embedding through a \textit{content-age aggregator} to generate a lifetime-aware news representation. The modular design allows this component to be seamlessly integrated into an existing \textit{news encoder}.
    \item \textbf{Lifetime-aware user modeling} constructs a user representation using a \textit{candidate-aware lifetime attention} mechanism, which assigns higher importance to clicked news that share similar topics with the candidate news. This encourages the model to focus on clicked news that are not only semantically relevant, but also aligned in terms of temporal interest, as measured by user-topic lifetime. As a result, this strategy enables the \textit{user encoder} to better capture both the user's semantic and temporal preferences toward the candidate news.
    \item \textbf{Freshness-guided interest matching} refines the final interest matching score via a \textit{freshness-guided interest refinement} module, which explicitly adjusts the user-news matching score based on how fresh the candidate news is for the user at the time of recommendation (i.e., prediction time).
\end{itemize}

\noindent
These three components enable {\method} to integrate \textit{implicit} (repre-sentation-level) and \textit{explicit} (matching-level) lifetime-aware interest matching. In addition, {\method} is \textit{orthogonally} compatible with existing neural news recommendation methods, making it broadly applicable across different base models. 
In the following subsections, we elaborate on each component of the framework in detail.

\subsection{Age-aware News Modeling}\label{sec:news-modeling}

\noindent
To precisely represent the time-sensitive nature of news, {\method} introduces a age-aware news modeling, which comprises of three modules: \textit{news encoder}, \textit{user-topic lifetime-aware age encoder,} and \textit{content-age aggregator}. These modules enrich news representations with both content semantics and temporal dynamics relative to user-topic lifetime.

\vspace{1mm}
\noindent
\textbf{(1) News encoder.}
We adopt a \textit{plug-in} design where any existing news encoder (e.g., NPA~\cite{wu2019npa}, NRMS~\cite{wu2019nrms}, NAML~\cite{wu2019naml}, CNE-SUE~\cite{mao2021cnesue}, or CROWN~\cite{ko2025crown}) can serve as a content encoder. 
Given the content attributes $D_n$, the content encoder extracts a semantic representation $\mathbf{h}_n^d \in \mathbb{R}^{d_d}$. 
This design allows {\method} to be flexibly applied across different news encoding methods without requiring structural modifications.

\vspace{1mm}
\noindent
\textbf{(2) User-Topic lifetime-aware age encoder.}
To represent how early or late a news article is consumed relative to the user's typical interest duration for the topic, 
we introduce a \textit{user-topic lifetime-aware age encoder}.
This module encodes the relative age of each news article as a function of its age and the user-topic lifetime.

Let $a$ denote the age of a news article (i.e., time since publish to click or prediction), 
and $L$ denote the user-topic lifetime, both measured in seconds. 
Since these values can span several orders of magnitude (e.g., from 300 to 864{,}000 seconds), 
we apply \textit{logarithmic bucketization}. The bucket index $b \in \{0, 1, ..., B-1\}$ is computed as:
\begin{align}
    b = \min\left( \left\lfloor \log(\max(a, 1)) \cdot \frac{B}{\log(\tau)} \right\rfloor, B - 1 \right)
    \label{eq:bucketization}
\end{align}
where $B$ is the total number of buckets, and $\tau$ is a scaling constant (e.g., $\tau=864{,}000$ seconds for a 10-day range). For each bucket ID of age $\mathbf{b}_a$ and lifetime $\mathbf{b}_l$, a low-dimensional vector (e.g., 200D), denoted as $\mathbf{e}_a$ and $\mathbf{e}_l$, is randomly initialized and learned during training. 
These vectors are concatenated and forwarded to a dense layer with non-linear activation to obtain the final user-topic lifetime-aware age representation:
\begin{align}
    \mathbf{h}_n^l = \tanh\left( \mathbf{W}_n \cdot [\mathbf{e}_a \oplus \mathbf{e}_l] + \mathbf{b}_n \right)
    \label{eq:age-embedding}
\end{align}

\noindent
where $\mathbf{W}_n$ and $\mathbf{b}_n$ are trainable parameters. The resulting embedding $\mathbf{h}_n^l \in \mathbb{R}^{d_l}$ represents the position of the news article within the user's topic-specific validity window: for clicked news, it reflects how early or late the article was clicked; for candidate news, it indicates how old the article is relative to the user-topic lifetime.

\vspace{1mm}
\noindent
\textbf{(3) Content-Age aggregator.}
To construct the final news representation, we combine the semantic embedding $\mathbf{h}_n^d$ and the user-topic lifetime-aware age embedding $\mathbf{h}_n^l$ using a fusion mechanism. 
In this paper, we adopt a simple yet effective concatenation strategy:
\begingroup
\setlength{\abovedisplayskip}{0.7em}
\setlength{\belowdisplayskip}{0.7em}
\setlength{\abovedisplayshortskip}{0.7em}
\setlength{\belowdisplayshortskip}{0.7em}
\begin{equation}
    \mathbf{h}_n^{dl} = \text{Concat}(\mathbf{h}_n^d, \mathbf{h}_n^l)
    \label{eq:concat}
\end{equation}
\endgroup
\noindent
The final news embedding $\mathbf{h}_n^{dl}$ carries different meanings depending on whether it represents a clicked or candidate news in terms of temporal interest. 
For clicked news, it indicates \textit{when the article was still valid to the user}. For candidate news, it shows \textit{whether the article is still valid at the time of recommendation}. 
\noindent
In summary, our method captures the \textit{relative age of news articles based on the user-topic lifetime} and combines it with the content embedding to generate a more informative representation. 
This enriched representation provides a \textit{crucial signal for precisely understanding the user's interest} in time-sensitive recommendation scenarios.

\subsection{Lifetime-aware User Modeling}\label{sec:user-modeling}

While some existing methods~\cite{qi2022caum, wang2018dkn, zhu2019dan, qi2021kim} incorporate the candidate news when constructing user representations, they mainly focus on modeling semantic relevance using the candidate content (e.g., title, body) as a clue. However, they do not consider whether \textit{the clicked news still remains valid to the user} given the user-topic interest duration. 
We argue that different clicked news should contribute differently not only based on their semantic relevance, 
but also based on whether they still represent valid interests for that topic at the time of recommendation.
To address this, we propose a \textit{candidate-aware lifetime attention} mechanism, which leverages the topic embedding of the candidate news as a \textit{query} to attend to clicked news that are semantically and temporally aligned with the candidate. 
Since topic acts as a compressed representation of content and serves as a basis for modeling user-topic lifetime, such a design enables our model to jointly align \textit{semantic and temporal preferences} when generating a user representation.

\vspace{1mm}
\noindent
\textbf{(1) Candidate-aware lifetime attention.}
Given a set of clicked news embeddings $\{\mathbf{h}_1^{dl}, ..., \mathbf{h}_N^{dl}\}$, where each $\mathbf{h}_i^{dl} \in \mathbb{R}^{d}$ is the content-age aggregated embedding of the $i$-th clicked news, we assign different importance scores to each clicked news based on its topic relevance to the candidate news.

To do so, we first compute the topic embedding of the candidate news $\mathbf{t}_C$ and the topic embeddings of the clicked news $\{\mathbf{t}_1, ..., \mathbf{t}_N\}$. Each topic embedding is obtained by combining the information of the news topic and sub-topic through a linear projection:
\begingroup
\setlength{\abovedisplayskip}{0.7em}
\setlength{\belowdisplayskip}{0.7em}
\setlength{\abovedisplayshortskip}{0.7em}
\setlength{\belowdisplayshortskip}{0.7em}
\begin{align}
\mathbf{t}_i = (\mathbf{c}_i \oplus \mathbf{s}_i) \cdot \mathbf{W}_{\text{t}} + \mathbf{b}_{\text{t}}, \quad
\mathbf{t}_C = (\mathbf{c}_C \oplus \mathbf{s}_C) \cdot \mathbf{W}_{\text{t}} + \mathbf{b}_{\text{t}}
\label{eq:topic-embedding}
\end{align}
\endgroup
where $\oplus$ denotes concatenation, $\mathbf{c}_i$ and $\mathbf{s}_i$ are the topic and sub-topic embeddings of the $i$-th clicked news, and $\mathbf{W}_{\text{t}}, \mathbf{b}_{\text{t}}$ are trainable parameters shared across clicked and candidate news.

We then compute attention weights $\{\alpha_1, ..., \alpha_N\}$ between the candidate topic embedding $\mathbf{t}_C$ and clicked topic embeddings $\mathbf{t}_i$ as:
\begin{align}
\alpha_i = \frac{\exp(\mathbf{t}_C^{\top} \mathbf{t}_i)}{\sum_{j=1}^{N} \exp(\mathbf{t}_C^{\top} \mathbf{t}_j)}
\label{eq:topic-attention}
\end{align}

\noindent
These attention scores are then used to reweight each clicked news embedding through a gated residual connection~\cite{chen2019gated, bresson2017residual}, which preserves the original information while adaptively incorporating attention weights that are more beneficial for user modeling.
Formally, the gated reweighting is defined as:
\begin{align}
\mathbf{h}_i^a = g_i \odot (\alpha_i \cdot \mathbf{h}_i^{dl}) + (1 - g_i) \odot \mathbf{h}_i^{dl}, \quad
g_i = \sigma(\mathbf{W}_g \cdot \mathbf{h}_i^{dl})
\label{eq:attention-gate}
\end{align}
where $\sigma$ denotes the sigmoid function, $\odot$ is element-wise multiplication, and $\mathbf{W}_g$ is a trainable weight matrix.

As a result, we obtain candidate-aware clicked news representations ${\mathbf{h}_1^a, ..., \mathbf{h}_N^a}$, where each embedding is adjusted to reflect semantic relevance and temporal interest to the candidate news.

\vspace{1mm}
\noindent
\textbf{(2) User encoder.}
Rather than proposing a new aggregation scheme, we design our framework to be \textit{model-agnostic} with respect to the user encoder. That is, the candidate-aware clicked news embeddings $\{\mathbf{h}_1^a, ..., \mathbf{h}_N^a\}$ can be seamlessly used as input to any existing user encoder, such as attention-based~\cite{wu2019nrms, wu2019npa, wu2019naml, qi2021hierec, li2022miner}, GRU-based~\cite{an2019lstur, okura2017embedding, zhu2019dan}, or graph-based encoders~\cite{mao2022digat, mao2021cnesue, yang2023glory, ko2025crown, ge2020graph}.

Our attention mechanism does not perform aggregation by itself but simply reweights the clicked news based on candidate-topic relevance. Therefore, it is model-agnostic and compatible with a wide range of existing user encoders. 

\subsection{Freshness-guided Interest Matching}\label{sec:interest-matching}

Most prior studies compute the interest matching score between the user and the candidate news solely based on their latent representations. However, these approaches \textit{fail to explicitly consider whether the candidate news is still valid for the user} at the time of recommendation.

\vspace{1mm}
\noindent
\textbf{(1) User-news interest matching.}
Given a user embedding $\mathbf{u}$ and a candidate news embedding $\mathbf{h}_c^{dl}$, we first compute the base interest matching score as their dot product:
\begin{align}
S_{\text{base}} = \mathbf{u}^{\top} \mathbf{h}_c^{dl}
\end{align}

\vspace{1mm}
\noindent
\textbf{(2) Freshness-guided interest refinement.}
To address the limitation of ignoring temporal interest, we propose a \textit{freshness-guided interest refinement} method that explicitly adjusts the matching score between a user and a candidate news article by considering the candidate’s relative freshness with respect to the user's topic-specific interest duration.

To incorporate the temporal validity of the candidate news, we define a freshness $F_c$ as the difference between the user-topic lifetime and the candidate news age (i.e., remaining lifetime). Intuitively, if a news article has expired (i.e., $F_c < 0$), it is less likely to be useful to the user. 
Inspired by the consideration of remaining lifetime in LANCER~\cite{bae2023lancer}, we adopt a scaled sigmoid function following prior work~\cite{bae2023lancer}, to adjust the matching score:
\begin{align}
f(F_c) =
\begin{cases}
\frac{1}{1 + \exp(-\alpha \cdot F_c)} & \text{if } F_c \geq 0 \\
\beta \cdot \frac{1}{1 + \exp(-\alpha \cdot F_c)} & \text{if } F_c < 0
\end{cases}
\end{align}
where $\alpha, \beta \in (0, 1)$ are scaling and penalty factors, respectively. The final freshness-guided interest score is computed as:
\begin{align}
S = S_{\text{base}} \cdot f(F_c)
\end{align}

\noindent
This approach ensures that news articles with insufficient freshness are down-weighted in the final ranking, while still respecting their base relevance. Compared to LANCER~\cite{bae2023lancer}, our method differs in two key aspects: (i) \textit{we compute freshness using fine-grained user-topic lifetime instead of a fixed global lifetime}, and (ii) \textit{we introduce a penalty factor to more effectively penalize expired news}. 
Ultimately, {\method} captures fine-grained user interests while ensuring that the recommended news remains temporally valid for the user.

\subsection{Model Training}\label{sec:proposed-training}

Based on the learned user and news representations,
{\method} is trained to predict whether a target user will click on a candidate news article.
The model parameters are optimized by minimizing the loss between the predicted and actual click labels.

\vspace{3mm}
\noindent
\textbf{Click prediction.}
Given a target user and a candidate news article,
we compute the predicted click score based on the dot product between their corresponding representations.
Note that we use the title embedding of the candidate news, i.e., $\mathbf{r}^n_{(T)}$, since users typically make click decisions based on the title they see~\cite{wu2019npa, mao2021cnesue, wu2019nrms, an2019lstur, wu2019naml, wu2019tanr, qi2021hierec, liu2020hypernews, wu2022feedrec, son2025rating}.
Formally, the predicted click score is calculated as:
\vspace{2mm}
\begin{align}
\hat{y}_{u,n} = \mathbf{r}^u \cdot \mathbf{r}^n_{(T)}
\end{align}
\vspace{2mm}
where $\mathbf{r}^u$ and $\mathbf{r}^n_{(T)}$ denote the final user and news title embeddings, respectively. To train the model, we adopt the negative sampling strategy widely used in prior work~\cite{mao2021cnesue, wang2018dkn, kim2022cast, qi2021kim, qi2021pprec, wu2019naml, wu2019nrms, kim2024polardsn, ko2022not, wu2020user, liu2020kred}.
Specifically, for each clicked news article (positive example),
we randomly sample $M$ unclicked news articles presented in the same impression as negative examples.
We then optimize the model so that the score of each positive news is ranked higher than those of the corresponding negatives. The final objective function, i.e., the click prediction loss $\mathcal{L}_P$, is defined as:
\vspace{2mm}
\begin{align}
    \mathcal{L}_P = -\sum_{i\in D} \log\left(\frac{exp(\hat{y}^+_{i})}{exp(\hat{y}^+_{i}) + \sum^M_{j=1} exp(\hat{y}^-_{i,j})}\right)
    \label{eq:click-loss}
\end{align}
\vspace{2mm}
where $D$ is the set of positive examples, 
$\hat{y}^+_{i}$ is the $i$-th positive example, 
and $\hat{y}^-_{i,j}$ is the $j$-th negative example for $\hat{y}^+_{i}$.

\vspace{1mm}
\noindent
We also analyze the time and space complexity of {\method}, as detailed in Section~\ref{sec:appendix-complexity}.

\section{Experimental Validation}\label{sec:eval}
In this section, we comprehensively evaluate {\method} by answering the following evaluation questions: 

\begin{itemize}[leftmargin=10pt]
    \item \textbf{EQ1} (\textbf{Overall Performance}): To what extent does {\method} improve the accuracy of state-of-the-art methods in personalized news recommendation?
    \item \textbf{EQ2} (\textbf{Ablation Study}): Are our strategies effective in improving performance individually and collectively?
    \item \textbf{EQ3} (\textbf{Impact of Lifetime Definition}): How do the different definitions of lifetime (fixed / topic-wise / user-topic) affect the performance of {\method}?
    \item \textbf{EQ4} (\textbf{Sensitivity Analysis}): How sensitive is {\method} to hyperparameters in freshness-guided weighting?
    \item \textbf{EQ5} (\textbf{Case Study}): How does {\method} provide explainable recommendations from a lifetime perspective?
\end{itemize}

\begin{table}[t]
\centering
\caption{Statistics of news article datasets.}
\vspace{-2mm}
\label{table:datasets}
\small
\setlength\tabcolsep{5pt}
\def\arraystretch{1.0} 
\begin{tabular}{l|rr}
\toprule
 
Dataset & \textbf{MIND-small}~\cite{wu2020mind} & \textbf{Adressa-1week}~\cite{gulla2017adressa} \\

\midrule
\textit{\# of users}  & 94,057 & 601,215 \\
\textit{\# of news articles}  & 65,238 & 73,844  \\
\textit{\# of clicks} &  347,727 & 2,107,312 \\
\textit{\# of categories} &  18 (270) & 108 (151)  \\
\midrule
\textit{\# of words-per-title} &  11.67 & 6.63 \\
\textit{\# of words-per-content} &  41.01 & 552.15 \\

\bottomrule
\end{tabular}
\end{table}

\subsection{Experimental Setup}\label{sec:eval-setup}

\noindent
\textbf{Datasets and baselines.}
Following prior works~\cite{bae2023lancer, gong2022tcar, yang2023glory, yi2021efficient}, we evaluate {\method} on two real-world datasets, \textbf{MIND-small} (\textbf{MIND} hereafter)~\cite{wu2020mind} and \textbf{Adressa-1week} (\textbf{Adressa} hereafter)~\cite{gulla2017adressa}.
Table~\ref{table:datasets} shows the statistics of these datasets.
We compare {\method} against five state-of-the-art news recommendation models: NRMS~\cite{wu2019nrms}, NAML~\cite{wu2019naml}, LSTUR~\cite{an2019lstur}, CNE-SUE~\cite{mao2021cnesue}, and CROWN~\cite{ko2025crown}. We use the official source codes for CNE-SUE and CROWN and adopt the open implementations available in the Microsoft NewsRec repository for NRMS, NAML, and LSTUR.
All models use the same 300-dimensional pre-trained word embeddings from GloVe~\cite{pennington2014glove} to ensure a fair comparison.
Since the publish and click times are \textit{not available} in MIND, following LANCER~\cite{bae2023lancer}, we treat the first impression time of a news article as its publish time, and the impression time of a clicked article for a user as her click time.

\vspace{1mm}
\noindent
\textbf{Evaluation protocol.}
Following previous works~\cite{li2022miner, ko2025crown, mao2022digat, yang2023glory, wu2019naml, qi2021pprec, ge2020graph, yu2022tiny}, we evaluate the top-\textit{n} recommendation accuracy by using four standard metrics: AUC (area under the ROC curve), MRR (mean reciprocal rank), and nDCG@5/10 (normalized discounted cumulative gain at rank 5 and 10).
We measure AUC, MRR, nDCG@5, and nDCG@10 on the test set at the epoch with the highest validation AUC, and report their averages over five runs.







\vspace{1mm}
\noindent
\textbf{Implementation details.}
We implement {\method} using PyTorch 1.12.1 on Ubuntu 20.04. All experiments are conducted on a machine with Intel i7-9700k CPU, 64GB memory, and NVIDIA RTX 2080 Ti GPU (CUDA 11.3, cuDNN 8.2.1).
The batch size is set to 32 for both datasets to maximize GPU utilization, where we use 4 negative samples per positive sample ($M=4$), following~\cite{wu2019nrms, an2019lstur, mao2021cnesue, ko2025crown, wu2019naml}.
We empirically set the scaling and penalty factors $\alpha$ and $\beta$ to 0.3 and 0.3, respectively, based on validation performance.
The implementation details, including all hyperparameters, are provided at {\codeurl}.

\begin{table*}[t]
\centering
\caption{News recommendation accuracy: {\method} consistently enhances \textit{all} competitors across \textit{all} metrics in a model-agnostic manner. The improvements are statistically significant under \textit{t}-tests with 95\% confidence (p-values $<$ 0.005).}\label{table:eval-eq1}
\vspace{-3mm}
\setlength\tabcolsep{4.5pt}
\footnotesize
\def\arraystretch{0.95}
\begin{tabular}{cc||ccc|ccc|ccc|ccc|ccc}
\toprule
& & \multicolumn{3}{c|}{NRMS~\cite{wu2019nrms}} & \multicolumn{3}{c|}{NAML~\cite{wu2019naml}} & \multicolumn{3}{c|}{LSTUR~\cite{an2019lstur}} & \multicolumn{3}{c|}{CNE-SUE~\cite{mao2021cnesue}} & \multicolumn{3}{c}{CROWN~\cite{ko2025crown}} \\
\cmidrule(lr){3-5}
\cmidrule(lr){6-8}
\cmidrule(lr){9-11}
\cmidrule(lr){12-14}
\cmidrule(lr){15-17}
& Metric & Base & \textbf{LIME} & Gain & Base & \textbf{LIME} & Gain & Base & \textbf{LIME} & Gain & Base & \textbf{LIME} & Gain & Base & \textbf{LIME} & Gain \\

\midrule
\multirow{4}{*}{\rotatebox{90}{\textbf{Adressa}}}
& AUC       & 0.7041 & \textbf{0.7933} & \textcolor{blue}{+12.67\%} & 0.7501 & \textbf{0.8316} & \textcolor{blue}{+10.87\%} & 0.7605 & \textbf{0.8347} & \textcolor{blue}{+9.76\%} & 0.7942 & \textbf{0.8627} & \textcolor{blue}{+8.62\%} & 0.8413 & \textbf{0.8774} & \textcolor{blue}{+4.29\%} \\
& MRR       & 0.4149 & \textbf{0.5185} & \textcolor{blue}{+24.97\%} & 0.4344 & \textbf{0.5312} & \textcolor{blue}{+22.28\%} & 0.5037 & \textbf{0.6025} & \textcolor{blue}{+19.61\%} & 0.4939 & \textbf{0.6055} & \textcolor{blue}{+22.59\%} & 0.5526 & \textbf{0.6534} & \textcolor{blue}{+18.24\%} \\
& nDCG@5    & 0.4046 & \textbf{0.5177} & \textcolor{blue}{+27.95\%} & 0.4435 & \textbf{0.5466} & \textcolor{blue}{+23.25\%} & 0.5038 & \textbf{0.6128} & \textcolor{blue}{+21.64\%} & 0.5196 & \textbf{0.6357} & \textcolor{blue}{+22.34\%} & 0.5865 & \textbf{0.6823} & \textcolor{blue}{+16.33\%} \\
& nDCG@10   & 0.4677 & \textbf{0.5835} & \textcolor{blue}{+24.76\%} & 0.5182 & \textbf{0.6005} & \textcolor{blue}{+15.88\%} & 0.5564 & \textbf{0.6606} & \textcolor{blue}{+18.73\%} & 0.5706 & \textbf{0.6736} & \textcolor{blue}{+18.05\%} & 0.6282 & \textbf{0.7123} & \textcolor{blue}{+13.39\%} \\
\midrule
\multirow{4}{*}{\rotatebox{90}{\textbf{MIND}}}
& AUC       & 0.6625 & \textbf{0.6672} & \textcolor{blue}{+0.71\%} & 0.6631 & \textbf{0.6694} & \textcolor{blue}{+0.95\%} & 0.6614 & \textbf{0.6675} & \textcolor{blue}{+0.92\%} & 0.6749 & \textbf{0.6816} & \textcolor{blue}{+0.99\%} & 0.6821 & \textbf{0.6884} & \textcolor{blue}{+0.92\%} \\
& MRR       & 0.3060 & \textbf{0.3241} & \textcolor{blue}{+5.92\%} & 0.3125 & \textbf{0.3282} & \textcolor{blue}{+5.02\%} & 0.3119 & \textbf{0.3224} & \textcolor{blue}{+3.37\%} & 0.3212 & \textbf{0.3346} & \textcolor{blue}{+4.17\%} & 0.3360 & \textbf{0.3422} & \textcolor{blue}{+1.85\%} \\
& nDCG@5    & 0.3432 & \textbf{0.3564} & \textcolor{blue}{+3.85\%} & 0.3437 & \textbf{0.3583} & \textcolor{blue}{+4.25\%} & 0.3428 & \textbf{0.3553} & \textcolor{blue}{+3.65\%} & 0.3558 & \textbf{0.3683} & \textcolor{blue}{+3.51\%} & 0.3721 & \textbf{0.3839} & \textcolor{blue}{+3.17\%} \\
& nDCG@10   & 0.4126 & \textbf{0.4195} & \textcolor{blue}{+1.67\%} & 0.4106 & \textbf{0.4208} & \textcolor{blue}{+2.48\%} & 0.4042 & \textbf{0.4189} & \textcolor{blue}{+3.64\%} & 0.4185 & \textbf{0.4290} & \textcolor{blue}{+2.51\%} & 0.4314 & \textbf{0.4392} & \textcolor{blue}{+1.81\%} \\
\bottomrule
\end{tabular}
\end{table*}

\begin{table*}[t]
\centering
\caption{Ablation study of {\method}: the proposed strategies (S1–S3) significantly contribute to enhancing the accuracy of {\method}. \underline{Underline} indicates the better result when the same number of strategies are applied. Bold indicates the best result in all cases.}
\label{table:ablation}
\vspace{-2mm}
\setlength\tabcolsep{4.8pt}
\footnotesize
\def\arraystretch{0.95}
\begin{tabular}{cc||c|cc|cc|cc|cc|cc}
\toprule
& & \begin{tabular}[c]{@{}c@{}}Baseline\end{tabular}
& \multicolumn{8}{c}{\textbf{LIME}} \\
\cmidrule(lr){3-3}
\cmidrule(lr){4-13}
& Metric & NAML & S1 & Gain & S3 & Gain & S1+S2 & Gain & S1+S3 & Gain & S1+S2+S3 & Gain \\
\midrule
\multirow{4}{*}{\rotatebox{90}{\textbf{Adressa}}}
& AUC       & 0.7501 & \underline{0.7943} & \textcolor{blue}{+5.89\%} 
& 0.7781 & \textcolor{blue}{+3.73\%}
& 0.4716 & \textcolor{blue}{+7.88\%} & \underline{0.8225} & \textcolor{blue}{+9.65\%} & \textbf{0.8316} & \textcolor{blue}{+10.87\%} \\
& MRR       & 0.4344 & \underline{0.4923} & \textcolor{blue}{+13.33\%} 
& 0.4716 & \textcolor{blue}{+8.56\%}
& 0.5120 & \textcolor{blue}{+17.86\%} & \underline{0.5253} & \textcolor{blue}{+20.93\%} & \textbf{0.5312} & \textcolor{blue}{+22.28\%} \\
& nDCG@5    & 0.4435 & \underline{0.5020} & \textcolor{blue}{+13.19\%} 
& 0.4832 & \textcolor{blue}{+8.95\%}
& 0.5268 & \textcolor{blue}{+18.78\%} & \underline{0.5384} & \textcolor{blue}{+21.40\%} & \textbf{0.5466} & \textcolor{blue}{+23.25\%} \\
& nDCG@10   & 0.5182 & \underline{0.5694} & \textcolor{blue}{+9.88\%} 
& 0.5450 & \textcolor{blue}{+5.17\%}
& 0.5860 & \textcolor{blue}{+13.08\%} & \underline{0.5922} & \textcolor{blue}{+14.28\%} & \textbf{0.6005} & \textcolor{blue}{+15.88\%} \\
\midrule
\multirow{4}{*}{\rotatebox{90}{\textbf{MIND}}}
& AUC       & 0.6631 & \underline{0.6656} & \textcolor{blue}{+0.38\%} 
& 0.6652 & \textcolor{blue}{+0.32\%}
& \underline{0.6681} & \textcolor{blue}{+0.75\%} & 0.6675 & \textcolor{blue}{+0.66\%} & \textbf{0.6694} & \textcolor{blue}{+0.95\%} \\
& MRR       & 0.3125 & \underline{0.3187} & \textcolor{blue}{+1.98\%} 
& 0.3172 & \textcolor{blue}{+1.50\%}
& \underline{0.3263} & \textcolor{blue}{+4.42\%} & 0.3241 & \textcolor{blue}{+3.71\%} & \textbf{0.3282} & \textcolor{blue}{+5.02\%} \\
& nDCG@5    & 0.3437 & \underline{0.3490} & \textcolor{blue}{+1.54\%} 
& 0.3488 & \textcolor{blue}{+1.48\%}
& \underline{0.3545} & \textcolor{blue}{+3.14\%} & 0.3522 & \textcolor{blue}{+2.47\%} & \textbf{0.3583} & \textcolor{blue}{+4.25\%} \\
& nDCG@10   & 0.4106 & \underline{0.4148} & \textcolor{blue}{+1.02\%} 
& 0.4140 & \textcolor{blue}{+0.83\%}
& \underline{0.4182} & \textcolor{blue}{+1.85\%} & 0.4172 & \textcolor{blue}{+1.61\%} & \textbf{0.4208} & \textcolor{blue}{+2.48\%} \\
\bottomrule
\end{tabular}
\end{table*}

\begin{table*}[t]
\centering
\caption{Impact of lifetime definitions across four base models. User-Topic lifetime consistently outperforms Fixed and Topic-wise variants across \textit{all} metrics. Bold and \underline{underline} indicate the best and second-best results, respectively.}
\label{table:eval-eq3}
\vspace{-2mm}
\footnotesize
\def\arraystretch{0.95}

\begin{minipage}[t]{\textwidth}
\centering
\vspace{1mm}

\begin{tabular}{cc||c|c@{\hskip 6pt}c|c@{\hskip 2pt}c|c@{\hskip 2pt}c||c|c@{\hskip 6pt}c|c@{\hskip 2pt}c|c@{\hskip 2pt}c}
\toprule

& Metric & NAML & \textbf{Fixed} & Gain & \textbf{Topic-wise}  & Gain & \textbf{User-Topic} & Gain & LSTUR & \textbf{Fixed} & Gain & \textbf{Topic-wise}  & Gain & \textbf{User-Topic} & Gain \\
\midrule
\multirow{4}{*}{\rotatebox{90}{\textbf{Adressa}}}
& AUC       & 0.7501 & 0.7768 & \textcolor{blue}{+3.6\%} & \underline{0.7994} & \textcolor{blue}{+6.6\%} & \textbf{0.8316} & \textcolor{blue}{+10.9\%} & 0.7605 & 0.7859 & \textcolor{blue}{+3.3\%} & \underline{0.7962} & \textcolor{blue}{+4.7\%} & \textbf{0.8347} & \textcolor{blue}{+9.8\%} \\
& MRR       & 0.4344 & 0.4525 & \textcolor{blue}{+4.2\%} & \underline{0.5109} & \textcolor{blue}{+17.6\%} & \textbf{0.5312} & \textcolor{blue}{+22.3\%} & 0.5037 & 0.5277 & \textcolor{blue}{+4.8\%} & \underline{0.5476} & \textcolor{blue}{+8.7\%} & \textbf{0.6025} & \textcolor{blue}{+19.6\%}\\
& nDCG@5    & 0.4435 & 0.4643 & \textcolor{blue}{+4.7\%} & \underline{0.5230} & \textcolor{blue}{+17.9\%} & \textbf{0.5466} & \textcolor{blue}{+23.3\%} & 0.5038 & 0.5287 & \textcolor{blue}{+4.9\%} & \underline{0.5552} & \textcolor{blue}{+10.2\%} & \textbf{0.6128} & \textcolor{blue}{+21.6\%} \\
& nDCG@10   & 0.5182 & 0.5317 & \textcolor{blue}{+2.6\%} & \underline{0.5793} & \textcolor{blue}{+11.8\%} & \textbf{0.6005} & \textcolor{blue}{+15.9\%} & 0.5564 & 0.5847 & \textcolor{blue}{+5.1\%} & \underline{0.6066} & \textcolor{blue}{+9.0\%} & \textbf{0.6606} & \textcolor{blue}{+18.7\%} \\
\midrule
\multirow{4}{*}{\rotatebox{90}{\textbf{MIND}}}
& AUC       & 0.6631 & 0.6636 & \textcolor{blue}{+0.1\%} & \underline{0.6672} & \textcolor{blue}{+0.6\%} & \textbf{0.6694} & \textcolor{blue}{+1.0\%} & 0.6614 & 0.6638 & \textcolor{blue}{+0.4\%} & \underline{0.6667} & \textcolor{blue}{+0.8\%} & \textbf{0.6675} & \textcolor{blue}{+0.9\%} \\
& MRR       & 0.3125 & 0.3109 & \textcolor{blue}{-0.5\%} & \underline{0.3215} & \textcolor{blue}{+2.9\%} & \textbf{0.3282} & \textcolor{blue}{+5.0\%} & 0.3119 & 0.3144 & \textcolor{blue}{+0.8\%} & \underline{0.3194} & \textcolor{blue}{+2.4\%} & \textbf{0.3224} & \textcolor{blue}{+3.4\%} \\
& nDCG@5    & 0.3437 & 0.3382 & \textcolor{blue}{-1.6\%} & \underline{0.3498} & \textcolor{blue}{+1.8\%} & \textbf{0.3583} & \textcolor{blue}{+4.3\%} & 0.3428 & 0.3486 & \textcolor{blue}{+1.7\%} & \underline{0.3520} & \textcolor{blue}{+2.7\%} & \textbf{0.3553} & \textcolor{blue}{+3.7\%} \\
& nDCG@10   & 0.4106 & 0.4091 & \textcolor{blue}{-0.4\%} & \underline{0.4162} & \textcolor{blue}{+1.4\%} & \textbf{0.4208} & \textcolor{blue}{+2.5\%} & 0.4042 & 0.4118 & \textcolor{blue}{+1.9\%} & \underline{0.4142} & \textcolor{blue}{+2.5\%} & \textbf{0.4189} & \textcolor{blue}{+3.6\%} \\
\bottomrule
\end{tabular}
\end{minipage}

\begin{minipage}[t]{\textwidth}
\centering
\vspace{1mm}

\begin{tabular}{cc||c|c@{\hskip 6pt}c|c@{\hskip 2pt}c|c@{\hskip 2pt}c||c|c@{\hskip 6pt}c|c@{\hskip 2pt}c|c@{\hskip 2pt}c}
\toprule

& Metric & CNE-SUE & \textbf{Fixed} & Gain & \textbf{Topic-wise} & Gain & \textbf{User-Topic} & Gain & CROWN & \textbf{Fixed} & Gain & \textbf{Topic-wise}  & Gain & \textbf{User-Topic} & Gain \\
\midrule
\multirow{4}{*}{\rotatebox{90}{\textbf{Adressa}}}
& AUC & 0.7942 & 0.8055 & \textcolor{blue}{+1.4\%} & \underline{0.8328} & \textcolor{blue}{+4.9\%} & \textbf{0.8627} & \textcolor{blue}{+8.6\%} & 0.8413 & 0.8468 & \textcolor{blue}{+0.7\%} & \underline{0.8612} & \textcolor{blue}{+2.4\%} & \textbf{0.8774} & \textcolor{blue}{+4.3\%} \\
& MRR & 0.4939 & 0.5482 & \textcolor{blue}{+10.9\%} & \underline{0.5825} & \textcolor{blue}{+17.9\%} & \textbf{0.6055} & \textcolor{blue}{+22.6\%} & 0.5526 & 0.5742 & \textcolor{blue}{+3.9\%} & \underline{0.6087} & \textcolor{blue}{+10.2\%} & \textbf{0.6534} & \textcolor{blue}{+18.2\%} \\
& nDCG@5    & 0.5196 & 0.5499 & \textcolor{blue}{+5.8\%} & \underline{0.5928} & \textcolor{blue}{+14.1\%} & \textbf{0.6357} & \textcolor{blue}{+22.3\%} & 0.5865 & 0.6061 & \textcolor{blue}{+3.3\%} & \underline{0.6568} & \textcolor{blue}{+12.0\%} & \textbf{0.6823} & \textcolor{blue}{+16.3\%} \\
& nDCG@10   & 0.5706 & 0.6152 & \textcolor{blue}{+7.8\%} & \underline{0.6406} & \textcolor{blue}{+12.3\%} & \textbf{0.6736} & \textcolor{blue}{+18.1\%} & 0.6282 & 0.6485 & \textcolor{blue}{+3.2\%} & \underline{0.6729} & \textcolor{blue}{+7.1\%} & \textbf{0.7123} & \textcolor{blue}{+13.4\%} \\
\midrule
\multirow{4}{*}{\rotatebox{90}{\textbf{MIND}}}
& AUC       & 0.6749 & 0.6776 & \textcolor{blue}{+0.4\%} & \underline{0.6790} & \textcolor{blue}{+0.6\%} & \textbf{0.6816} & \textcolor{blue}{+1.0\%} & 0.6821 & 0.6831 & \textcolor{blue}{+0.2\%} & \underline{0.6852} & \textcolor{blue}{+0.5\%} & \textbf{0.6884} & \textcolor{blue}{+0.9\%} \\
& MRR       & 0.3212 & 0.3241 & \textcolor{blue}{+0.9\%} & \underline{0.3293} & \textcolor{blue}{+2.5\%} & \textbf{0.3346} & \textcolor{blue}{+4.2\%} & 0.3360 & 0.3392 & \textcolor{blue}{+1.0\%} & \underline{0.3401} & \textcolor{blue}{+1.2\%} & \textbf{0.3422} & \textcolor{blue}{+1.9\%} \\
& nDCG@5    & 0.3558 & 0.3568 & \textcolor{blue}{+0.3\%} & \underline{0.3628} & \textcolor{blue}{+2.0\%} & \textbf{0.3683} & \textcolor{blue}{+3.5\%} & 0.3721 & 0.3748 & \textcolor{blue}{+0.7\%} & \underline{0.3796} & \textcolor{blue}{+2.0\%} & \textbf{0.3839} & \textcolor{blue}{+3.2\%} \\
& nDCG@10   & 0.4185 & 0.4198 & \textcolor{blue}{+0.3\%} & \underline{0.4226} & \textcolor{blue}{+1.0\%} & \textbf{0.4290} & \textcolor{blue}{+2.5\%} & 0.4314 & 0.4342 & \textcolor{blue}{+0.7\%} & \underline{0.4380} & \textcolor{blue}{+1.5\%} & \textbf{0.4392} & \textcolor{blue}{+1.8\%} \\
\bottomrule
\end{tabular}
\end{minipage}

\vspace{4mm}

\end{table*}

\subsection{Overall Performance (EQ1)}\label{sec:eval-result-eq1}
Table~\ref{table:eval-eq1} shows the recommendation accuracies of {\method} and all competing methods on the two datasets. The results demonstrate that {\method} \textit{consistently} improves \textit{all} state-of-the-art in \textit{all} metrics on \textit{both} datasets.
Specifically, {\method} achieves substantial gains when applied to NRMS~\cite{wu2019nrms}, improving AUC, MRR, nDCG@5, and nDCG@10 by up to 12.67\%, 24.97\%, 27.95\%, and 24.76\% in \textbf{Adressa}, respectively.
Even on the strongest competitor (CROWN~\cite{ko2025crown}), {\method} still achieves notable improvements of 18.24\% and 16.33\% in MRR and nDCG@5, respectively.
We note that these improvements of {\method} over all competing models are remarkable, given that they are achieved \textit{without modifying} any model structure and through a simple yet effective modular integration.
We also conducted the \textit{t}-tests with a 95\% confidence level and verified that the improvement of {\method} over all competing methods are statistically significant (i.e., all $p$-values are below 0.005).
Consequently, these results demonstrate that {\method} accurately captures a user's interest in a temporal aspect by precisely leveraging the age of news in the user's click history (C1), and effectively modeling the varying lifetime across topics and users (C2), through the following strategies:
(S1) user-topic lifetime-aware age representation, 
(S2) candidate-aware lifetime attention, and 
(S3) freshness-guided interest refinement.



\subsection{Ablation Study (EQ2)}\label{sec:eval-result-eq2}

We verify the effectiveness of \textit{our proposed strategies} in {\method}. We consider all possible combinations of the following three key strategies: (S1) \textit{User-Topic lifetime-aware age representation}, (S2) \textit{Candidate-aware lifetime attention}, and (S3) \textit{Freshness-guided interest refinement}. We note that S2 operates on the news embeddings generated by S1; therefore, it cannot be applied independently.

Table~\ref{table:ablation} shows the results of the ablation study.
Overall, each of our proposed strategies consistently contributes to improving the accuracy of {\method}.  
In particular, when \textit{all} three strategies (S1, S2, S3) are applied together, {\method} achieves the \textit{best} performance, improving AUC, MRR, nDCG@5, and nDCG@10 by up to 10.87\%, 22.28\%, 23.25\%, and 15.88\%, respectively, in \textbf{Adressa}.  
These results demonstrate that:  
(1) comprehending users’ interests in a temporal perspective is crucial for accurate personalized news recommendation, and  
(2) the proposed strategies in {\method} successfully address this challenge by incorporating temporal signals into news modeling, user modeling, and interest matching.

Looking more closely, when a single strategy is applied, {\method} with S1 consistently outperforms {\method} with S3, showing up to 13.33\% gain in MRR.  
This result suggests that precisely representing a news article by understanding its age with respect to the user-topic lifetime is most critical in the time-sensitive nature of news recommendation, thereby verifying the effect of S1, as discussed in Section~\ref{sec:news-modeling}.
When two strategies are applied, S1+S2 and S1+S3 achieve the highest accuracy on \textbf{MIND} and \textbf{Adressa}, respectively, with S1+S3 showing up to 21.40\% gain in \textbf{Adressa}.
These results imply that depending on the characteristics of each news dataset, either enhancing user modeling or refining interest matching can play a more important role.
In both cases, the results verify the effectiveness of S2 and S3, as described in Sections~\ref{sec:user-modeling} and~\ref{sec:interest-matching}, respectively.
Finally, when \textit{all} three strategies are applied, {\method} achieves the \textit{best} result in all cases. 
This result demonstrates that accurately understanding \textit{age}, \textit{lifetime}, and \textit{freshness} of news article, and integrating these temporal signals together into the news modeling, user modeling, and interest matching, is essential for better capturing users' interests in a temporal aspect and achieving more accurate news recommendation.

\subsection{Impact of Lifetime Definition (EQ3)}\label{sec:eval-result-eq3}

We verify the impact of the lifetime definition on the accuracy of {\method}.  
We compare three variants of {\method} that adopt different lifetime definitions across four competitors in two datasets.  
(1) \textbf{Fixed} performs all three strategies of {\method} using \textit{a fixed lifetime} value (i.e., \textit{36 hours}~\cite{bae2023lancer}).  
(2) \textbf{Topic-wise} and \textbf{User-Topic} perform the proposed strategies based on our definitions of \textit{topic-wise lifetime} and \textit{user-topic lifetime}, respectively, where \textbf{User-Topic} is our finally selected definition.  
Table~\ref{table:eval-eq3} shows the results. \textbf{User-Topic} consistently achieves the best accuracy across \textit{all} metrics and two datasets.  
These results demonstrate that: our lifetime definition (1) not only aligns with the real-world click behaviors, (2) but also effectively reduces prediction errors by capturing users' interests accurately in a temporal perspective, as we claimed in Section~\ref{sec:motivation}.

More specifically, the \textbf{Fixed} lifetime achieves accuracy comparable to or slightly better than the baselines alone, which indicates that \textit{a single fixed value} fails to fully exploit the potential of time-sensitive nature of news lifetime.  
When lifetime is defined in a \textbf{Topic-wise}, it achieves higher accuracy than \textbf{Fixed}, with up to 17.9\% gain in \textbf{Adressa}.
These results verify that the lifetime of news articles varies across topics, which helps better understand both news content and user preferences.
Consequently, we demonstrate that carefully incorporating the two proposed lifetime definitions into {\method} effectively unleashes the potential of news lifetime.

\begin{figure}[t]
\centering
\hspace*{-6mm}
\begin{tabular}{cc}
    \centering
    \includegraphics[width=0.46\linewidth]{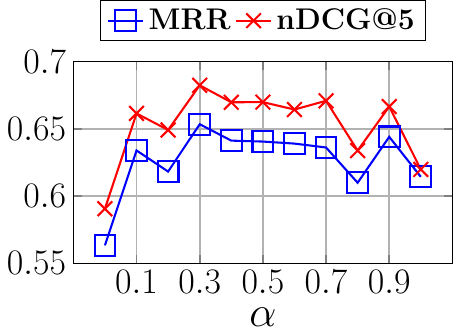} &
     \raisebox{-1.6mm}{\includegraphics[width=0.46\linewidth]{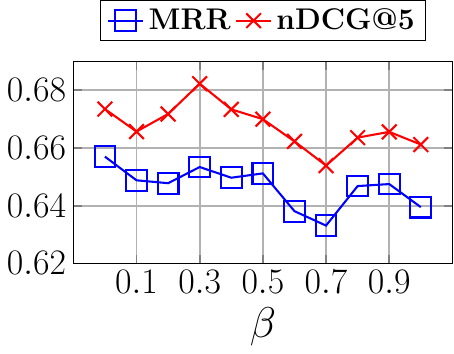}} \\
    (a) Scaling factor  & (b) Penalty factor
\end{tabular}
\vspace{-3mm}
\caption{Sensitivity to freshness-guided interest matching hyperparameters $\alpha$ and $\beta$.}\label{fig:eval-eq4}
\vspace{-4mm}
\end{figure}

\subsection{Sensitivity Analysis (EQ4)}\label{sec:eval-result-eq4}

We evaluate the impact of the two hyperparameters in the freshness-guided interest matching strategy, namely $\alpha$ and $\beta$, on the accuracy of {\method} built upon the CROWN model.
We vary $\alpha$ and $\beta$ from 0.0 to 1.0 in increment of 0.1 and measure the accuracy of {\method}.  
Figure~\ref{fig:eval-eq4} shows the results, where the \textit{x}-axis represents each control factor ($\alpha$ or $\beta$) and the \textit{y}-axis represents the accuracy (i.e., MRR and nDCG@5).
For $\alpha$, the accuracy of {\method} tends to increase from 0.0 (i.e., the freshness-guided interest matching is not used) and peaks in the range of $0.3 \leq \alpha \leq 0.7$, after which it gradually decreases toward 1.0.  
For $\beta$, the accuracy slightly drops from 0.0 (i.e., a strong penalty is applied when the candidate news is expired) to 0.1, and reaches its highest value at $\beta = 0.3$.  
After that, the accuracy gradually decreases as $\beta$ increases toward 1.0 (i.e., the penalty is not used).
These results demonstrate that the proposed freshness-guided interest matching provides a signal beneficial to recommend candidate news that remains valid to the user. Furthermore, they verify that both $\alpha$ and $\beta$ serve as effective regularization factors to prevent overfitting to either outdated or irrelevant candidate news.


\subsection{Case Study (EQ5)}\label{sec:eval-result-eq5}

\begin{figure}[t]
\centering
\hspace*{-3mm}
\begin{tabular}{c}
    \centering
    \includegraphics[width=1.00\linewidth]{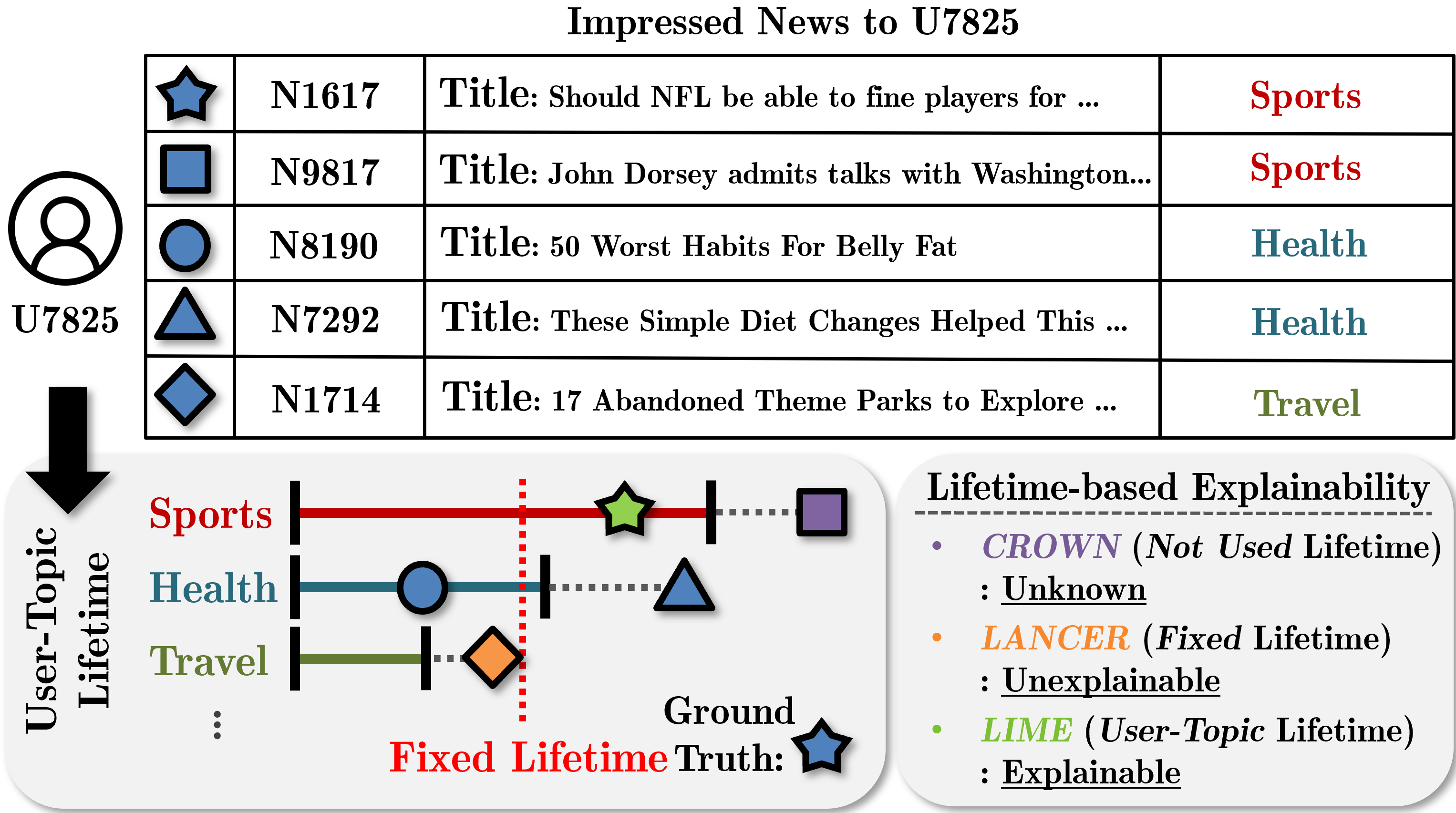}
\end{tabular}
\vspace{-3mm}
\caption{Case study for explainable recommendations from a lifetime perspective.}\label{fig:eval-eq5}
\vspace{-4mm}
\end{figure}

We conduct a case study to further highlight the effectiveness of {\method}.  
We compare {\method} in plug-in/out form with CROWN, and indirectly compare it with LANCER by applying a \textit{fixed lifetime} to {\method}.  
Figure~\ref{fig:eval-eq5} shows a sampled impression (i.e., candidate news) where user \textit{U7825} has previously clicked several news articles.
With {\method}, we can capture the user-topic lifetime for \textit{U7825}, which is visualized in the figure.  
Each colored shape placed above the user-topic lifetime timeline indicates that the corresponding method predicted the user would click that candidate news.  
For instance, CROWN predicts that the user would click the purple square news (i.e., \textit{N9817}) among the five impressed news articles.
In contrast, {\method} correctly recommends the ground-truth candidate news (i.e., \textit{N1617}) by identifying which news remains valid to \textit{U7825} within the user-topic lifetime. 
This allows for a \textit{lifetime-based explainability} in which the recommendation can be reasonably justified from a temporal perspective.
However, CROWN does not consider lifetime in the recommendation. Although the content may be interesting to the user, it may no longer be valid at the time of recommendation. As a result, it recommends a outdated news (i.e., \textit{N9817}).
Furthermore, when using \textit{fixed lifetime} (i.e., indirectly representing LANCER), the model considers the expiration of news but fails to adaptively capture the variations across different topics and users. Consequently, it recommends a candidate news (i.e., \textit{N1714}) that the user has already lost interest in.
These results demonstrate that {\method} better understands the time-sensitive nature of news and recommends candidate news that is still valid and interesting to individual users, with explainability from a temporal perspective.

\subsection{Complexity Analysis}\label{sec:appendix-complexity}
In this section, we analyze the space and time complexity of {\method}.
\noindent
\textbf{Space complexity.} {\method} consists of (1) a news encoder, (2) a user encoder, (3) a click predictor, and (4) a freshness-guided refinement layer. 
The parameter size of a news encoder requires $O(d^2)$ for standard encoding and $O(2d^2)$ for our additional content-age aggregation, where $d$ is the embedding dimensionality. 
For the age encoder, {\method} introduces a bucket embedding table of size $B \times d$, which is negligible since $B \ll |N|$. 
The parameter size of a user encoder is $O(|U| \cdot d + |N| \cdot d)$ for storing user and news embeddings, plus $O(d^2)$ for attention parameters. 
The click predictor requires $O(d)$ for a dot product, and the freshness-guided refinement module introduces two scalar parameters $\alpha$ and $\beta$, and a small number of dense layers (e.g., $O(d^2)$). 
Therefore, since $B, \alpha, \beta \ll |U|, |N|$, the overall space complexity of {\method} is $O((|U| + |N| + d) \cdot d)$, which remains linear in the number of users and news articles. 
Although {\method} includes additional modules (e.g., age encoder and freshness refinement), their overhead is much smaller than the common embedding space (e.g., $O(d^2 + 2d^2) \ll O(|U| \cdot d + |N| \cdot d)$), making it comparable to existing methods.

\noindent
\textbf{Time complexity.} The computational overhead of {\method} consists of (1) news representation, (2) user representation, and (3) interest score computation. 
The age-aware news representation adds $O(1)$ bucket lookup and $O(d^2)$ for a projection layer. 
The candidate-aware attention for user modeling involves dot products and gating across $N$ clicked news, resulting in $O(N \cdot d^2)$. 
The interest score refinement requires $O(1)$ for freshness calculation and $O(d)$ for reweighting. 
Hence, the overall time complexity of {\method} is $O(N \cdot d^2)$, which is comparable to standard attention-based models and acceptable in practice, as shown in our experiments.


\section{Related Work}\label{sec:related}
In this section, we review existing works for news recommendation.
Many neural news recommendation methods~\cite{wu2019npa, wu2019nrms, wu2019naml, an2019lstur, wu2019tanr, wang2020fim, liu2020hypernews, qi2021hierec, mao2021cnesue, mao2022digat, chen2023tccm, yang2023glory, li2022miner, ko2025crown} aim to improve user interest modeling via enhanced news and user representations. For example, NPA~\cite{wu2019npa} applies user-specific attention to highlight informative words/news for personalized reading preference.
NRMS~\cite{wu2019nrms} adopts multi-head self-attention to capture the semantic relatedness among clicked news. NAML~\cite{wu2019naml} integrates titles, bodies, and categories via multi-view attention to learn richer news representations. LSTUR~\cite{an2019lstur} models short-/long-term interests via ID embeddings and GRU. FIM~\cite{wang2020fim} performs fine-grained matching between semantic segments of candidate and browsed news at multiple levels. CNE-SUE~\cite{mao2021cnesue} jointly encodes title and body via collaborative attention, and leverages GCNs for structural user modeling. CROWN~\cite{ko2025crown} employs category-guided intent disentanglement, consistency-aware news encoding, and GNN-based user modeling to address intent diversity, post-read preference variance, and cold-start users.
these methods often overlook the time-sensitive nature of news~\cite{wu2020mind, liu2020hypernews, wu2023personalized, qi2021pprec, chen2023tccm, bae2023lancer, ko2023khan}. To address this, recent studies have begun incorporating time-related attributes such as age and lifetime~\cite{liu2020hypernews, qi2021pprec, chen2023tccm, bae2023lancer}.
For example, HyperNews~\cite{liu2020hypernews} models candidate news age via a timeliness module, jointly predicting click and reading time. PP-Rec~\cite{qi2021pprec} uses recency-aware popularity and a popularity-aware user encoder to address cold-start and diversity. TCCM~\cite{chen2023tccm} introduces a causal framework that considers both content and temporal effects in modeling user-news interactions. LANCER~\cite{bae2023lancer} first defines news lifetime and penalizes expired news under limited competition.
However, these approaches have two key limitations. First, they focus only on the \textit{age of candidate news}, overlooking the informative signal in the age of clicked news when modeling user interests. Second, the existing definition of lifetime~\cite{bae2023lancer} assumes a globally fixed value, ignoring its potential variation across topics and users.
To our best knowledge, {\method} is the first to incorporate (i) clicked news age for user modeling and (ii) user-topic lifetime definitions, enabling more accurate recommendation in time-sensitive news scenarios.

\section{Conclusion}\label{sec:conclusion}
In this paper, we identify two key challenges in personalized news recommendation: (C1) accurately leveraging the age of clicked news for user modeling, and (C2) modeling varying news lifetime across topics and users. To address these challenges, we propose a model-agnostic framework, named as {\method}, which incorporates time-sensitive attributes into both representation and matching stages. Specifically, {\method} employs (1) the user-topic lifetime-aware age representation for (C1), (2) the candidate-aware lifetime attention for (C2), and (3) the freshness-guided interest refinement to ensure the temporal validity of recommended news. Furthermore, we define new lifetime concepts (topic-wise and user-topic lifetime), and validate their alignment with real user behaviors and their potential to better capture the time-sensitive nature of news recommendation. Experimental results on two real-world datasets show that (1) \textit{(Accuracy)} {\method} significantly improves the accuracy of state-of-the-art models, and (2) \textit{(Effectiveness)} each proposed strategy contributes to better modeling of lifetime-aware interest matching in the time-sensitive news domain.


\begin{acks}\label{sec:ack}
The work of Sang-Wook Kim was supported by Institute of Information \& Communications Technology Planning \& Evaluation (IITP) grant funded by the Korea government (MSIT) (RS-2022-00155586, RS-2020-II201373).
The work of Yunyong Ko was supported by the National Research Foundation of Korea (NRF) grant funded by the Korea government (MSIT) (RS-2024-00459301).
\end{acks}

\section*{GenAI Usage Disclosure}\label{sec:genai_disclosure}
In accordance with the ACM Authorship Policy, we disclose that generative AI tools (e.g., ChatGPT) were used exclusively to assist in editing and polishing the author-written content, including minor grammar correction, phrasing edits, and word-level autocorrection.

\bibliographystyle{ACM-Reference-Format}
\bibliography{bibliography}


\begin{thebibliography}{40}


\ifx \showCODEN    \undefined \def \showCODEN     #1{\unskip}     \fi
\ifx \showISBNx    \undefined \def \showISBNx     #1{\unskip}     \fi
\ifx \showISBNxiii \undefined \def \showISBNxiii  #1{\unskip}     \fi
\ifx \showISSN     \undefined \def \showISSN      #1{\unskip}     \fi
\ifx \showLCCN     \undefined \def \showLCCN      #1{\unskip}     \fi
\ifx \shownote     \undefined \def \shownote      #1{#1}          \fi
\ifx \showarticletitle \undefined \def \showarticletitle #1{#1}   \fi
\ifx \showURL      \undefined \def \showURL       {\relax}        \fi
\providecommand\bibfield[2]{#2}
\providecommand\bibinfo[2]{#2}
\providecommand\natexlab[1]{#1}
\providecommand\showeprint[2][]{arXiv:#2}

\bibitem[An et~al\mbox{.}(2019)]%
        {an2019lstur}
\bibfield{author}{\bibinfo{person}{Mingxiao An}, \bibinfo{person}{Fangzhao Wu}, \bibinfo{person}{Chuhan Wu}, \bibinfo{person}{Kun Zhang}, \bibinfo{person}{Zheng Liu}, {and} \bibinfo{person}{Xing Xie}.} \bibinfo{year}{2019}\natexlab{}.
\newblock \showarticletitle{Neural news recommendation with long-and short-term user representations}. In \bibinfo{booktitle}{\emph{Proceedings of the 57th Annual Meeting of the Association for Computational Linguistics}}. \bibinfo{pages}{336--345}.
\newblock


\bibitem[Bae et~al\mbox{.}(2023)]%
        {bae2023lancer}
\bibfield{author}{\bibinfo{person}{Hong-Kyun Bae}, \bibinfo{person}{Jeewon Ahn}, \bibinfo{person}{Dongwon Lee}, {and} \bibinfo{person}{Sang-Wook Kim}.} \bibinfo{year}{2023}\natexlab{}.
\newblock \showarticletitle{LANCER: A Lifetime-Aware News Recommender System}. In \bibinfo{booktitle}{\emph{Proceedings of the AAAI Conference on Artificial Intelligence}}, Vol.~\bibinfo{volume}{37}. \bibinfo{pages}{4141--4148}.
\newblock


\bibitem[Bresson and Laurent(2017)]%
        {bresson2017residual}
\bibfield{author}{\bibinfo{person}{Xavier Bresson} {and} \bibinfo{person}{Thomas Laurent}.} \bibinfo{year}{2017}\natexlab{}.
\newblock \showarticletitle{Residual gated graph convnets}.
\newblock \bibinfo{journal}{\emph{arXiv preprint arXiv:1711.07553}} (\bibinfo{year}{2017}).
\newblock


\bibitem[Chen et~al\mbox{.}(2019)]%
        {chen2019gated}
\bibfield{author}{\bibinfo{person}{Cen Chen}, \bibinfo{person}{Kenli Li}, \bibinfo{person}{Sin~G Teo}, \bibinfo{person}{Xiaofeng Zou}, \bibinfo{person}{Kang Wang}, \bibinfo{person}{Jie Wang}, {and} \bibinfo{person}{Zeng Zeng}.} \bibinfo{year}{2019}\natexlab{}.
\newblock \showarticletitle{Gated residual recurrent graph neural networks for traffic prediction}. In \bibinfo{booktitle}{\emph{Proceedings of the AAAI conference on artificial intelligence}}, Vol.~\bibinfo{volume}{33}. \bibinfo{pages}{485--492}.
\newblock


\bibitem[Chen et~al\mbox{.}(2023)]%
        {chen2023tccm}
\bibfield{author}{\bibinfo{person}{Yewang Chen}, \bibinfo{person}{Weiyao Ye}, \bibinfo{person}{Guipeng Xv}, \bibinfo{person}{Chen Lin}, {and} \bibinfo{person}{Xiaomin Zhu}.} \bibinfo{year}{2023}\natexlab{}.
\newblock \showarticletitle{Tccm: time and content-aware causal model for unbiased news recommendation}. In \bibinfo{booktitle}{\emph{Proceedings of the 32nd ACM International Conference on Information and Knowledge Management}}. \bibinfo{pages}{3778--3782}.
\newblock


\bibitem[Cho et~al\mbox{.}(2021)]%
        {cho2021overlooked}
\bibfield{author}{\bibinfo{person}{Sungmin Cho}, \bibinfo{person}{Hongjun Lim}, \bibinfo{person}{Keunchan Park}, \bibinfo{person}{Sungjoo Yoo}, {and} \bibinfo{person}{Eunhyeok Park}.} \bibinfo{year}{2021}\natexlab{}.
\newblock \showarticletitle{On the overlooked significance of underutilized contextual features in recent news recommendation models}.
\newblock \bibinfo{journal}{\emph{arXiv preprint arXiv:2112.14370}} (\bibinfo{year}{2021}).
\newblock


\bibitem[Ge et~al\mbox{.}(2020)]%
        {ge2020graph}
\bibfield{author}{\bibinfo{person}{Suyu Ge}, \bibinfo{person}{Chuhan Wu}, \bibinfo{person}{Fangzhao Wu}, \bibinfo{person}{Tao Qi}, {and} \bibinfo{person}{Yongfeng Huang}.} \bibinfo{year}{2020}\natexlab{}.
\newblock \showarticletitle{Graph enhanced representation learning for news recommendation}. In \bibinfo{booktitle}{\emph{Proceedings of The Web Conference 2020}}. \bibinfo{pages}{2863--2869}.
\newblock


\bibitem[Gong and Zhu(2022)]%
        {gong2022tcar}
\bibfield{author}{\bibinfo{person}{Shansan Gong} {and} \bibinfo{person}{Kenny~Q. Zhu}.} \bibinfo{year}{2022}\natexlab{}.
\newblock \showarticletitle{Positive, Negative and Neutral: Modeling Implicit Feedback in Session-Based News Recommendation}. In \bibinfo{booktitle}{\emph{Proceedings of the 45th International ACM SIGIR Conference on Research and Development in Information Retrieval}} (Madrid, Spain) \emph{(\bibinfo{series}{SIGIR '22})}. \bibinfo{address}{New York, USA}, \bibinfo{pages}{1185–1195}.
\newblock
\showISBNx{9781450387323}


\bibitem[Gulla et~al\mbox{.}(2017)]%
        {gulla2017adressa}
\bibfield{author}{\bibinfo{person}{Jon~Atle Gulla}, \bibinfo{person}{Lemei Zhang}, \bibinfo{person}{Peng Liu}, \bibinfo{person}{{\"O}zlem {\"O}zg{\"o}bek}, {and} \bibinfo{person}{Xiaomeng Su}.} \bibinfo{year}{2017}\natexlab{}.
\newblock \showarticletitle{The adressa dataset for news recommendation}. In \bibinfo{booktitle}{\emph{Proceedings of the international conference on web intelligence}}. \bibinfo{pages}{1042--1048}.
\newblock


\bibitem[Kim et~al\mbox{.}(2024)]%
        {kim2024polardsn}
\bibfield{author}{\bibinfo{person}{Min-Jeong Kim}, \bibinfo{person}{Yeon-Chang Lee}, {and} \bibinfo{person}{Sang-Wook Kim}.} \bibinfo{year}{2024}\natexlab{}.
\newblock \showarticletitle{PolarDSN: An Inductive Approach to Learning the Evolution of Network Polarization in Dynamic Signed Networks}. In \bibinfo{booktitle}{\emph{Proceedings of the 33rd ACM International Conference on Information and Knowledge Management}}. \bibinfo{pages}{1099--1109}.
\newblock


\bibitem[Kim et~al\mbox{.}(2022)]%
        {kim2022cast}
\bibfield{author}{\bibinfo{person}{Taeho Kim}, \bibinfo{person}{Yungi Kim}, \bibinfo{person}{Yeon-Chang Lee}, \bibinfo{person}{Won-Yong Shin}, {and} \bibinfo{person}{Sang-Wook Kim}.} \bibinfo{year}{2022}\natexlab{}.
\newblock \showarticletitle{Is It Enough Just Looking at the Title? Leveraging Body Text To Enrich Title Words Towards Accurate News Recommendation}. In \bibinfo{booktitle}{\emph{Proceedings of the CIKM}}. \bibinfo{pages}{4138--4142}.
\newblock


\bibitem[Ko et~al\mbox{.}(2022)]%
        {ko2022not}
\bibfield{author}{\bibinfo{person}{Yunyong Ko}, \bibinfo{person}{Dongwon Lee}, {and} \bibinfo{person}{Sang-Wook Kim}.} \bibinfo{year}{2022}\natexlab{}.
\newblock \showarticletitle{Not all layers are equal: A layer-wise adaptive approach toward large-scale dnn training}. In \bibinfo{booktitle}{\emph{Proceedings of the ACM Web Conference 2022}}. \bibinfo{pages}{1851--1859}.
\newblock


\bibitem[Ko et~al\mbox{.}(2023)]%
        {ko2023khan}
\bibfield{author}{\bibinfo{person}{Yunyong Ko}, \bibinfo{person}{Seongeun Ryu}, \bibinfo{person}{Soeun Han}, \bibinfo{person}{Youngseung Jeon}, \bibinfo{person}{Jaehoon Kim}, \bibinfo{person}{Sohyun Park}, \bibinfo{person}{Kyungsik Han}, \bibinfo{person}{Hanghang Tong}, {and} \bibinfo{person}{Sang-Wook Kim}.} \bibinfo{year}{2023}\natexlab{}.
\newblock \showarticletitle{KHAN: Knowledge-Aware Hierarchical Attention Networks for Accurate Political Stance Prediction}. In \bibinfo{booktitle}{\emph{Proceedings of the ACM Web Conference 2023}}. \bibinfo{pages}{1572--1583}.
\newblock


\bibitem[Ko et~al\mbox{.}(2025)]%
        {ko2025crown}
\bibfield{author}{\bibinfo{person}{Yunyong Ko}, \bibinfo{person}{Seongeun Ryu}, {and} \bibinfo{person}{Sang-Wook Kim}.} \bibinfo{year}{2025}\natexlab{}.
\newblock \showarticletitle{CROWN: A Novel Approach to Comprehending Users' Preferences for Accurate Personalized News Recommendation}. In \bibinfo{booktitle}{\emph{Proceedings of the ACM on Web Conference 2025}}. \bibinfo{pages}{1911--1921}.
\newblock


\bibitem[Li et~al\mbox{.}(2022)]%
        {li2022miner}
\bibfield{author}{\bibinfo{person}{Jian Li}, \bibinfo{person}{Jieming Zhu}, \bibinfo{person}{Qiwei Bi}, \bibinfo{person}{Guohao Cai}, \bibinfo{person}{Lifeng Shang}, \bibinfo{person}{Zhenhua Dong}, \bibinfo{person}{Xin Jiang}, {and} \bibinfo{person}{Qun Liu}.} \bibinfo{year}{2022}\natexlab{}.
\newblock \showarticletitle{MINER: multi-interest matching network for news recommendation}. In \bibinfo{booktitle}{\emph{Findings of the ACL}}. \bibinfo{pages}{343--352}.
\newblock


\bibitem[Liu et~al\mbox{.}(2020a)]%
        {liu2020kred}
\bibfield{author}{\bibinfo{person}{Danyang Liu}, \bibinfo{person}{Jianxun Lian}, \bibinfo{person}{Shiyin Wang}, \bibinfo{person}{Ying Qiao}, \bibinfo{person}{Jiun-Hung Chen}, \bibinfo{person}{Guangzhong Sun}, {and} \bibinfo{person}{Xing Xie}.} \bibinfo{year}{2020}\natexlab{a}.
\newblock \showarticletitle{KRED: Knowledge-aware document representation for news recommendations}. In \bibinfo{booktitle}{\emph{Proceedings of the 14th ACM Conference on Recommender Systems}}. \bibinfo{pages}{200--209}.
\newblock


\bibitem[Liu et~al\mbox{.}(2020b)]%
        {liu2020hypernews}
\bibfield{author}{\bibinfo{person}{Rui Liu}, \bibinfo{person}{Huilin Peng}, \bibinfo{person}{Yong Chen}, {and} \bibinfo{person}{Dell Zhang}.} \bibinfo{year}{2020}\natexlab{b}.
\newblock \showarticletitle{HyperNews: Simultaneous News Recommendation and Active-Time Prediction via a Double-Task Deep Neural Network.}. In \bibinfo{booktitle}{\emph{IJCAI}}, Vol.~\bibinfo{volume}{20}. \bibinfo{pages}{3487--3493}.
\newblock


\bibitem[Mao et~al\mbox{.}(2022)]%
        {mao2022digat}
\bibfield{author}{\bibinfo{person}{Zhiming Mao}, \bibinfo{person}{Jian Li}, \bibinfo{person}{Hongru Wang}, \bibinfo{person}{Xingshan Zeng}, {and} \bibinfo{person}{Kam-Fai Wong}.} \bibinfo{year}{2022}\natexlab{}.
\newblock \showarticletitle{DIGAT: Modeling News Recommendation with Dual-Graph Interaction}. In \bibinfo{booktitle}{\emph{Findings of the Association for Computational Linguistics: EMNLP 2022}}. \bibinfo{pages}{6595--6607}.
\newblock


\bibitem[Mao et~al\mbox{.}(2021)]%
        {mao2021cnesue}
\bibfield{author}{\bibinfo{person}{Zhiming Mao}, \bibinfo{person}{Xingshan Zeng}, {and} \bibinfo{person}{Kam-Fai Wong}.} \bibinfo{year}{2021}\natexlab{}.
\newblock \showarticletitle{Neural News Recommendation with Collaborative News Encoding and Structural User Encoding}. In \bibinfo{booktitle}{\emph{Findings of the Association for Computational Linguistics: EMNLP 2021}}. \bibinfo{address}{Punta Cana, Dominican Republic}, \bibinfo{pages}{46--55}.
\newblock


\bibitem[Okura et~al\mbox{.}(2017)]%
        {okura2017embedding}
\bibfield{author}{\bibinfo{person}{Shumpei Okura}, \bibinfo{person}{Yukihiro Tagami}, \bibinfo{person}{Shingo Ono}, {and} \bibinfo{person}{Akira Tajima}.} \bibinfo{year}{2017}\natexlab{}.
\newblock \showarticletitle{Embedding-based news recommendation for millions of users}. In \bibinfo{booktitle}{\emph{Proceedings of the 23rd ACM SIGKDD international conference on knowledge discovery and data mining}}. \bibinfo{pages}{1933--1942}.
\newblock


\bibitem[Pennington et~al\mbox{.}(2014)]%
        {pennington2014glove}
\bibfield{author}{\bibinfo{person}{Jeffrey Pennington}, \bibinfo{person}{Richard Socher}, {and} \bibinfo{person}{Christopher~D Manning}.} \bibinfo{year}{2014}\natexlab{}.
\newblock \showarticletitle{Glove: Global vectors for word representation}. In \bibinfo{booktitle}{\emph{Proceedings of the 2014 conference on empirical methods in natural language processing (EMNLP)}}. \bibinfo{pages}{1532--1543}.
\newblock


\bibitem[Qi et~al\mbox{.}(2021a)]%
        {qi2021kim}
\bibfield{author}{\bibinfo{person}{Tao Qi}, \bibinfo{person}{Fangzhao Wu}, \bibinfo{person}{Chuhan Wu}, {and} \bibinfo{person}{Yongfeng Huang}.} \bibinfo{year}{2021}\natexlab{a}.
\newblock \showarticletitle{Personalized news recommendation with knowledge-aware interactive matching}. In \bibinfo{booktitle}{\emph{Proceedings of the 44th International ACM SIGIR Conference}}. \bibinfo{pages}{61--70}.
\newblock


\bibitem[Qi et~al\mbox{.}(2021b)]%
        {qi2021pprec}
\bibfield{author}{\bibinfo{person}{Tao Qi}, \bibinfo{person}{Fangzhao Wu}, \bibinfo{person}{Chuhan Wu}, {and} \bibinfo{person}{Yongfeng Huang}.} \bibinfo{year}{2021}\natexlab{b}.
\newblock \showarticletitle{PP-Rec: News Recommendation with Personalized User Interest and Time-aware News Popularity}. In \bibinfo{booktitle}{\emph{Proceedings of the 59th Annual Meeting of the Association for Computational Linguistics and the 11th International Joint Conference on Natural Language Processing (Volume 1: Long Papers)}}. \bibinfo{pages}{5457--5467}.
\newblock


\bibitem[Qi et~al\mbox{.}(2022)]%
        {qi2022caum}
\bibfield{author}{\bibinfo{person}{Tao Qi}, \bibinfo{person}{Fangzhao Wu}, \bibinfo{person}{Chuhan Wu}, {and} \bibinfo{person}{Yongfeng Huang}.} \bibinfo{year}{2022}\natexlab{}.
\newblock \showarticletitle{News recommendation with candidate-aware user modeling}. In \bibinfo{booktitle}{\emph{Proceedings of the 45th international ACM SIGIR conference on research and development in information retrieval}}. \bibinfo{pages}{1917--1921}.
\newblock


\bibitem[Qi et~al\mbox{.}(2021c)]%
        {qi2021hierec}
\bibfield{author}{\bibinfo{person}{Tao Qi}, \bibinfo{person}{Fangzhao Wu}, \bibinfo{person}{Chuhan Wu}, \bibinfo{person}{Peiru Yang}, \bibinfo{person}{Yang Yu}, \bibinfo{person}{Xing Xie}, {and} \bibinfo{person}{Yongfeng Huang}.} \bibinfo{year}{2021}\natexlab{c}.
\newblock \showarticletitle{HieRec: Hierarchical User Interest Modeling for Personalized News Recommendation}. In \bibinfo{booktitle}{\emph{Proceedings of the 59th Annual Meeting of the Association for Computational Linguistics and the 11th International Joint Conference on Natural Language Processing (Volume 1: Long Papers)}}. \bibinfo{pages}{5446--5456}.
\newblock


\bibitem[Son et~al\mbox{.}(2025)]%
        {son2025rating}
\bibfield{author}{\bibinfo{person}{Jiwon Son}, \bibinfo{person}{Hyunjoon Kim}, {and} \bibinfo{person}{Sang-Wook Kim}.} \bibinfo{year}{2025}\natexlab{}.
\newblock \showarticletitle{Rating-Aware Homogeneous Review Graphs and User Likes/Dislikes Differentiation for Effective Recommendations}. In \bibinfo{booktitle}{\emph{Proceedings of the 48th International ACM SIGIR Conference on Research and Development in Information Retrieval}}. \bibinfo{pages}{2070--2080}.
\newblock


\bibitem[Wang et~al\mbox{.}(2020)]%
        {wang2020fim}
\bibfield{author}{\bibinfo{person}{Heyuan Wang}, \bibinfo{person}{Fangzhao Wu}, \bibinfo{person}{Zheng Liu}, {and} \bibinfo{person}{Xing Xie}.} \bibinfo{year}{2020}\natexlab{}.
\newblock \showarticletitle{Fine-grained interest matching for neural news recommendation}. In \bibinfo{booktitle}{\emph{Proceedings of the 58th annual meeting of the association for computational linguistics}}. \bibinfo{pages}{836--845}.
\newblock


\bibitem[Wang et~al\mbox{.}(2018)]%
        {wang2018dkn}
\bibfield{author}{\bibinfo{person}{Hongwei Wang}, \bibinfo{person}{Fuzheng Zhang}, \bibinfo{person}{Xing Xie}, {and} \bibinfo{person}{Minyi Guo}.} \bibinfo{year}{2018}\natexlab{}.
\newblock \showarticletitle{DKN: Deep knowledge-aware network for news recommendation}. In \bibinfo{booktitle}{\emph{Proceedings of the 2018 world wide web conference}}. \bibinfo{pages}{1835--1844}.
\newblock


\bibitem[Wu et~al\mbox{.}(2019b)]%
        {wu2019naml}
\bibfield{author}{\bibinfo{person}{Chuhan Wu}, \bibinfo{person}{Fangzhao Wu}, \bibinfo{person}{Mingxiao An}, \bibinfo{person}{Jianqiang Huang}, \bibinfo{person}{Yongfeng Huang}, {and} \bibinfo{person}{Xing Xie}.} \bibinfo{year}{2019}\natexlab{b}.
\newblock \showarticletitle{Neural news recommendation with attentive multi-view learning}. In \bibinfo{booktitle}{\emph{Proceedings of the 28th IJCAI}}. \bibinfo{pages}{3863--3869}.
\newblock


\bibitem[Wu et~al\mbox{.}(2019c)]%
        {wu2019npa}
\bibfield{author}{\bibinfo{person}{Chuhan Wu}, \bibinfo{person}{Fangzhao Wu}, \bibinfo{person}{Mingxiao An}, \bibinfo{person}{Jianqiang Huang}, \bibinfo{person}{Yongfeng Huang}, {and} \bibinfo{person}{Xing Xie}.} \bibinfo{year}{2019}\natexlab{c}.
\newblock \showarticletitle{NPA: neural news recommendation with personalized attention}. In \bibinfo{booktitle}{\emph{Proceedings of the 25th ACM SIGKDD international conference on knowledge discovery \& data mining}}. \bibinfo{pages}{2576--2584}.
\newblock


\bibitem[Wu et~al\mbox{.}(2019a)]%
        {wu2019tanr}
\bibfield{author}{\bibinfo{person}{Chuhan Wu}, \bibinfo{person}{Fangzhao Wu}, \bibinfo{person}{Mingxiao An}, \bibinfo{person}{Yongfeng Huang}, {and} \bibinfo{person}{Xing Xie}.} \bibinfo{year}{2019}\natexlab{a}.
\newblock \showarticletitle{Neural news recommendation with topic-aware news representation}. In \bibinfo{booktitle}{\emph{Proceedings of the 57th ACL}}. \bibinfo{pages}{1154--1159}.
\newblock


\bibitem[Wu et~al\mbox{.}(2019d)]%
        {wu2019nrms}
\bibfield{author}{\bibinfo{person}{Chuhan Wu}, \bibinfo{person}{Fangzhao Wu}, \bibinfo{person}{Suyu Ge}, \bibinfo{person}{Tao Qi}, \bibinfo{person}{Yongfeng Huang}, {and} \bibinfo{person}{Xing Xie}.} \bibinfo{year}{2019}\natexlab{d}.
\newblock \showarticletitle{Neural news recommendation with multi-head self-attention}. In \bibinfo{booktitle}{\emph{Proceedings of the 2019 conference on empirical methods in natural language processing and the 9th international joint conference on natural language processing}}. \bibinfo{pages}{6389--6394}.
\newblock


\bibitem[Wu et~al\mbox{.}(2023)]%
        {wu2023personalized}
\bibfield{author}{\bibinfo{person}{Chuhan Wu}, \bibinfo{person}{Fangzhao Wu}, \bibinfo{person}{Yongfeng Huang}, {and} \bibinfo{person}{Xing Xie}.} \bibinfo{year}{2023}\natexlab{}.
\newblock \showarticletitle{Personalized news recommendation: Methods and challenges}.
\newblock \bibinfo{journal}{\emph{ACM Transactions on Information Systems}} \bibinfo{volume}{41}, \bibinfo{number}{1} (\bibinfo{year}{2023}), \bibinfo{pages}{1--50}.
\newblock


\bibitem[Wu et~al\mbox{.}(2020b)]%
        {wu2020user}
\bibfield{author}{\bibinfo{person}{Chuhan Wu}, \bibinfo{person}{Fangzhao Wu}, \bibinfo{person}{Tao Qi}, {and} \bibinfo{person}{Yongfeng Huang}.} \bibinfo{year}{2020}\natexlab{b}.
\newblock \showarticletitle{User Modeling with Click Preference and Reading Satisfaction for News Recommendation.}. In \bibinfo{booktitle}{\emph{IJCAI}}. \bibinfo{pages}{3023--3029}.
\newblock


\bibitem[Wu et~al\mbox{.}(2022)]%
        {wu2022feedrec}
\bibfield{author}{\bibinfo{person}{Chuhan Wu}, \bibinfo{person}{Fangzhao Wu}, \bibinfo{person}{Tao Qi}, \bibinfo{person}{Qi Liu}, \bibinfo{person}{Xuan Tian}, \bibinfo{person}{Jie Li}, \bibinfo{person}{Wei He}, \bibinfo{person}{Yongfeng Huang}, {and} \bibinfo{person}{Xing Xie}.} \bibinfo{year}{2022}\natexlab{}.
\newblock \showarticletitle{Feedrec: News feed recommendation with various user feedbacks}. In \bibinfo{booktitle}{\emph{Proceedings of the ACM Web Conference 2022}}. \bibinfo{pages}{2088--2097}.
\newblock


\bibitem[Wu et~al\mbox{.}(2020a)]%
        {wu2020mind}
\bibfield{author}{\bibinfo{person}{Fangzhao Wu}, \bibinfo{person}{Ying Qiao}, \bibinfo{person}{Jiun-Hung Chen}, \bibinfo{person}{Chuhan Wu}, \bibinfo{person}{Tao Qi}, \bibinfo{person}{Jianxun Lian}, \bibinfo{person}{Danyang Liu}, \bibinfo{person}{Xing Xie}, \bibinfo{person}{Jianfeng Gao}, \bibinfo{person}{Winnie Wu}, {et~al\mbox{.}}} \bibinfo{year}{2020}\natexlab{a}.
\newblock \showarticletitle{Mind: A large-scale dataset for news recommendation}. In \bibinfo{booktitle}{\emph{Proceedings of the 58th Annual Meeting of the Association for Computational Linguistics}}. \bibinfo{pages}{3597--3606}.
\newblock


\bibitem[Yang et~al\mbox{.}(2023)]%
        {yang2023glory}
\bibfield{author}{\bibinfo{person}{Boming Yang}, \bibinfo{person}{Dairui Liu}, \bibinfo{person}{Toyotaro Suzumura}, \bibinfo{person}{Ruihai Dong}, {and} \bibinfo{person}{Irene Li}.} \bibinfo{year}{2023}\natexlab{}.
\newblock \showarticletitle{Going Beyond Local: Global Graph-Enhanced Personalized News Recommendations}. In \bibinfo{booktitle}{\emph{Proceedings of the ACM Conference on Recommender Systems}}. \bibinfo{pages}{24--34}.
\newblock


\bibitem[Yi et~al\mbox{.}(2021)]%
        {yi2021efficient}
\bibfield{author}{\bibinfo{person}{Jingwei Yi}, \bibinfo{person}{Fangzhao Wu}, \bibinfo{person}{Chuhan Wu}, \bibinfo{person}{Ruixuan Liu}, \bibinfo{person}{Guangzhong Sun}, {and} \bibinfo{person}{Xing Xie}.} \bibinfo{year}{2021}\natexlab{}.
\newblock \showarticletitle{Efficient-FedRec: Efficient Federated Learning Framework for Privacy-Preserving News Recommendation}. In \bibinfo{booktitle}{\emph{Proceedings of the EMNLP}}. \bibinfo{pages}{2814--2824}.
\newblock


\bibitem[Yu et~al\mbox{.}(2022)]%
        {yu2022tiny}
\bibfield{author}{\bibinfo{person}{Yang Yu}, \bibinfo{person}{Fangzhao Wu}, \bibinfo{person}{Chuhan Wu}, \bibinfo{person}{Jingwei Yi}, {and} \bibinfo{person}{Qi Liu}.} \bibinfo{year}{2022}\natexlab{}.
\newblock \showarticletitle{Tiny-NewsRec: Effective and Efficient PLM-based News Recommendation}. In \bibinfo{booktitle}{\emph{Proceedings of the 2022 Conference on Empirical Methods in Natural Language Processing}}. \bibinfo{pages}{5478--5489}.
\newblock


\bibitem[Zhu et~al\mbox{.}(2019)]%
        {zhu2019dan}
\bibfield{author}{\bibinfo{person}{Qiannan Zhu}, \bibinfo{person}{Xiaofei Zhou}, \bibinfo{person}{Zeliang Song}, \bibinfo{person}{Jianlong Tan}, {and} \bibinfo{person}{Li Guo}.} \bibinfo{year}{2019}\natexlab{}.
\newblock \showarticletitle{Dan: Deep attention neural network for news recommendation}. In \bibinfo{booktitle}{\emph{Proceedings of the AAAI Conference on Artificial Intelligence}}, Vol.~\bibinfo{volume}{33}. \bibinfo{pages}{5973--5980}.
\newblock


\end{thebibliography}

\end{document}